\documentclass[useAMS,usenatbib]{mn2e}

\usepackage{graphicx,times}
\usepackage{hyperref}
\def\gtrsim{\mathrel{\hbox{\rlap{\hbox{\lower4pt\hbox{$\sim$}}}\hbox{$>$}}}}

\title[Calibrating the HFF cluster lenses]{Mass and magnification maps for the \textit{Hubble Space Telescope} Frontier Fields clusters:  implications for high redshift studies}
\author[J.\ Richard and the CATS team]{
 \parbox[h]{\textwidth}{
Johan Richard$^1$\thanks{E-mail:
johan.richard@univ-lyon1.fr},  Mathilde Jauzac$^{2,3}$, 
Marceau Limousin$^4$, Eric Jullo$^4$, Benjamin Cl\'ement$^5$, Harald Ebeling$^6$, Jean-Paul Kneib$^7$, Hakim Atek$^7$, Priya Natarajan$^8$,
Eiichi Egami$^5$, Rachael Livermore$^9$, Richard Bower$^3$}
\vspace{6pt}\\
$^{1}$CRAL, Observatoire de Lyon, Universit\'e Lyon 1, 9 Avenue Ch.\ Andr\'e, 69561 Saint Genis Laval Cedex, France\\
$^{2}$Astrophysics and Cosmology Research Unit, School of Mathematical Sciences, University of KwaZulu-Natal, Durban 4041, South Africa\\
$^{3}$Institute for Computational Cosmology, Durham University, South Road, Durham DH1 3LE, U.K.\\
$^{4}$Laboratoire d'Astrophysique de Marseille, CNRS- Universit\'e Aix-Marseille, 38 rue F. Joliot-Curie, 13388 Marseille Cedex 13, France\\
$^{5}$Steward Observatory, University of Arizona, 933 N. Cherry Avenue, Tucson, AZ 85721, USA\\
$^{6}$Institute for Astronomy, University of Hawaii, 2680 Woodlawn Drive, Honolulu, HI 96822, USA\\
$^{7}$ Laboratoire d'Astrophysique, Ecole Polytechnique F\'ed\'erale de Lausanne, Observatoire de Sauverny, CH-1290 Versoix, Switzerland\\
$^{8}$ Department of Astronomy, Yale University, 260 Whitney Avenue, New Haven, CT 06511, USA\\
$^{9}$ Department of Astronomy, the University of Texas at Austin, 2515 Speedway Stop C1400, Austin, TX 78712, USA \\
 }

\date{Draft version; for submission to MNRAS}

\begin{document}

\maketitle

\begin{abstract}
Extending over three \textit{Hubble Space Telescope} (HST) cycles, the Hubble Frontier Fields (HFF) initiative constitutes the largest commitment ever of HST time to the exploration of the distant Universe via gravitational lensing by massive galaxy clusters. Here, we present models of the mass distribution in the six  HFF cluster lenses, derived from a joint strong- and weak-lensing analysis anchored by a total of 88 multiple-image systems identified in existing HST data. The resulting maps of the projected mass distribution and of the gravitational magnification effectively calibrate the HFF clusters as gravitational telescopes. Allowing the computation of search areas in the source plane, these maps are provided to the community to facilitate the exploitation of forthcoming HFF data for quantitative studies of the gravitationally lensed population of background galaxies. Our models of the gravitational magnification afforded by the HFF clusters allow us to quantify the lensing-induced boost in sensitivity over blank-field observations and predict that galaxies at $z>10$ and as faint as m(AB)=32 will be detectable, up to 2 magnitudes fainter than the limit of the Hubble Ultra Deep Field.

\end{abstract}


\section{Gravitational lensing}

At the end of the 1980s the discovery of giant luminous arcs in clusters of galaxies \citep[e.g.,][]{1988A&A...191L..19S} and the realization that they can be explained by gravitational lensing (the bending of light by massive foreground mass concentrations) opened a new and powerful route to studying the distant Universe. Gravitational lensing can be understood as a geometrical mapping of the source plane onto the image plane, a mapping that depends on the surface mass distribution in the deflector and on the angular-diameter distances between the observer and the source, and the lens and the source, respectively \citep[see the review by][]{2011A&ARv..19...47K}. For extreme mass concentrations the mapping is non-linear, producing magnified and highly distorted multiple images of a single background source. Such multiple images have been successfully exploited since the early 1990s \citep[e.g.,][]{1996ApJ...471..643K} to constrain the detailed mass distribution in cluster cores as well as to probe  the mass distributions of samples of X- ray selected clusters out to the virial radius \citep[e.g.,][]{2005MNRAS.359..417S,2009MNRAS.395.1213E,2010MNRAS.404..325R,2012ApJ...749...97Z}. Importantly, lensing also enables the study of distant background galaxies that would be unobservable without the magnification provided by the cluster lens \citep[e.g.,][]{2001ApJ...560L.119E,2004ApJ...607..697K,2008ApJ...685..705R,2009ApJ...690.1764B,2013ApJ...762...32C}, and offers the tantalizing possibility of measuring the geometry of the Universe through the accurate determination of cosmological distances \citep{2004A&A...417L..33S,2010Sci...329..924J}.

The power of clusters as well-calibrated telescopes for studies of the distant Universe has become fully appreciated only in recent years. Highly active areas of research aiming to unveil the characteristics of individual galaxies at $1<z<5$ \citep[e.g.,][]{2007ApJ...654L..33S,2009MNRAS.400.1121S,2011MNRAS.413..643R,2012MNRAS.427..688L} or to constrain statistically the properties of the galaxy population at $z>5$ \citep[e.g.,][]{2011ApJ...737...90B} all benefit greatly from gravitational amplification of the respective background sources. State-of-the art lens-modeling techniques that combine strong-lensing constraints from large numbers of multiple-image systems with high-quality weak-lensing data can measure the mass in cluster cores (and thus the gravitational amplification along a given line of sight) to an accuracy of a few percent \citep[e.g.,][]{Bradac07,Bradac09,2007NJPh....9..447J,2009MNRAS.395.1319J}. 

The necessary robust and efficient identification of multiple-image systems requires both high angular resolution and color information. The unparalleled power of \textit{HST} for such studies is exemplified by the identification of 42 multiple-image systems in Advanced Camera for Surveys (ACS)  observations of the massive cluster Abell 1689 \citep{2005ApJ...621...53B}, 24 of them with measured redshifts from deep Keck and VLT spectroscopy \citep{2007ApJ...668..643L}.  The ability to obtain spectra for gravitationally amplified galaxies at high redshift is a critical advantage over similar work conducted on non-magnified galaxies in the field, and crucial for the exploration of the end of the Dark Ages, one of the most ambitious and timely quests of present-day astrophysics.  Although impressive progress has been made in this research area with the help of moderately deep observations of cluster lenses \citep[e.g.,][]{Postman2012}, a dedicated in-depth observational effort is needed if the scientific promise and potential of gravitational lensing by clusters is to be fully exploited.

This important  next step forward is now being taken in the form of the \textit{Hubble Frontier Fields} (HFF), a recent initiative launched by the Space Telescope Science Institute. As part of the preparations for these unprecedented observations of lensing clusters, five independent teams have analysed the existing imaging and spectroscopic data to provide the community with accurate mass models on each cluster. We describe in this paper the work performed by one of these groups, the CATS (Clusters As TelescopeS) team.

\section{The Hubble Frontier Fields}
The \textit{Hubble Space Telescope} Frontier Fields initiative, announced in the spring of 2013, devotes 140 orbits of \textit{HST} time to deep imaging observations of each of six carefully selected cluster lenses. As a compromise between depth and spectral coverage, each target field will be observed for 20 orbits in each of the F435W, F606W, and F814W filters (all ACS), as well as in the F105W, F125W, F140W, and F160W filters (all WFC3), reaching $m\approx 29$ (AB) uniformly in all passbands. The total commitment of 840 orbits of Director's Discretionary Time is spread out over three cycles, starting with Cycle 21, with two clusters being targeted per cycle.

The HFF clusters were selected by an expert team recruited from the extragalactic community, trying to balance scheduling and follow-up constraints with the primary goal of the project: to identify clusters of maximal lensing strength (high gravitational magnification over a large angular area) whose angular size is well matched to the ACS field of view. The resulting list of HFF clusters comprises, in order of observation\\[2mm]
\noindent
\begin{tabular}{lcl}
Name &   $z$   & reference \\ \hline
Abell 2744                               & 0.308 & \citet{1958ApJS....3..211A}\\
MACS\,J0416.1$-$2403   &0.396 & \citet{2012MNRAS.420.2120M}\\
MACS\,J0717.5$+$3745  & 0.545 & \citet{2007ApJ...661L..33E}\\
MACS\,J1149.5$+$2223  & 0.544 & \citet{2007ApJ...661L..33E}\\
Abell S1063                              & 0.348 & \citet{1989ApJS...70....1A}\\
Abell 370                                   & 0.375 & \citet{1958ApJS....3..211A}\\ \hline \\
\end{tabular}

More information on the HFF initiative, both scientific and technical, can be found on the \href{http://www.stsci.edu/hst/campaigns/frontier-fields/}{\textit{HST} Frontier Fields homepage\footnote{\url{http://www.stsci.edu/hst/campaigns/frontier-fields/}}}.

All six HFF clusters have previously been targeted with \textit{HST}. These observations were instrumental in the selection of the respective clusters for the HFF project as highly efficient gravitational lenses. Unlike A370, the first cluster lens to be discovered \citep{1988A&A...191L..19S,Richard2010a}, most of the remaining HFF targets are much more recent discoveries made by the Massive Cluster Survey \citep[MACS,][]{2001ApJ...553..668E}. As a result, many of the existing \textit{HST} observations of these fields were obtained only in the past few years and could thus take advantage of \textit{HST}'s current state-of-the-art instruments. Further details are provided in Table~\ref{tab:hst-obs} which lists all imaging observations performed with broad-band filters on ACS and WFC3. 

\begin{figure*}
\includegraphics[width=0.3\textwidth,clip,trim=50mm 79mm 50mm 65mm]{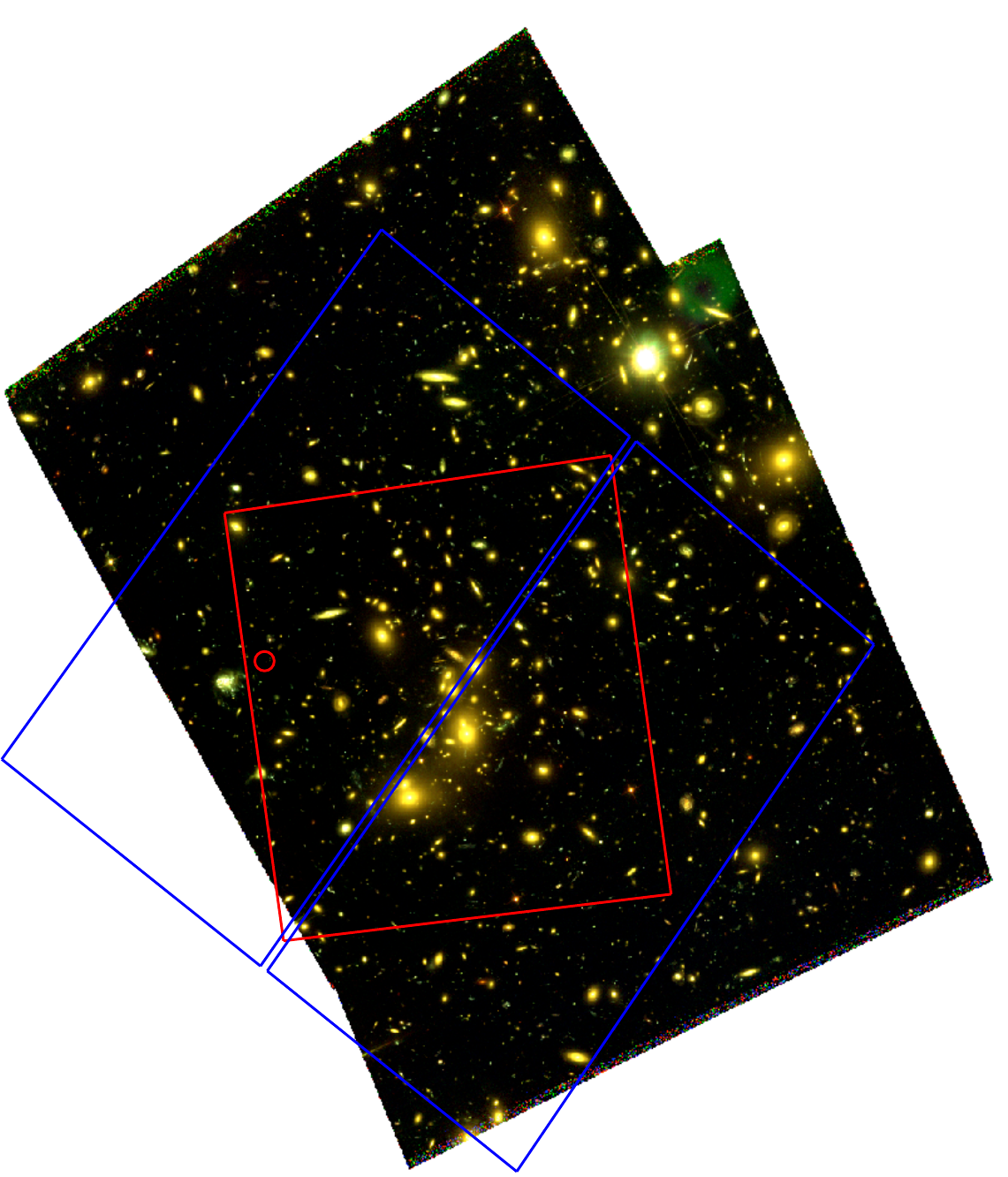}
\includegraphics[width=0.33\textwidth,clip,trim=60mm 100mm 60mm 80mm]{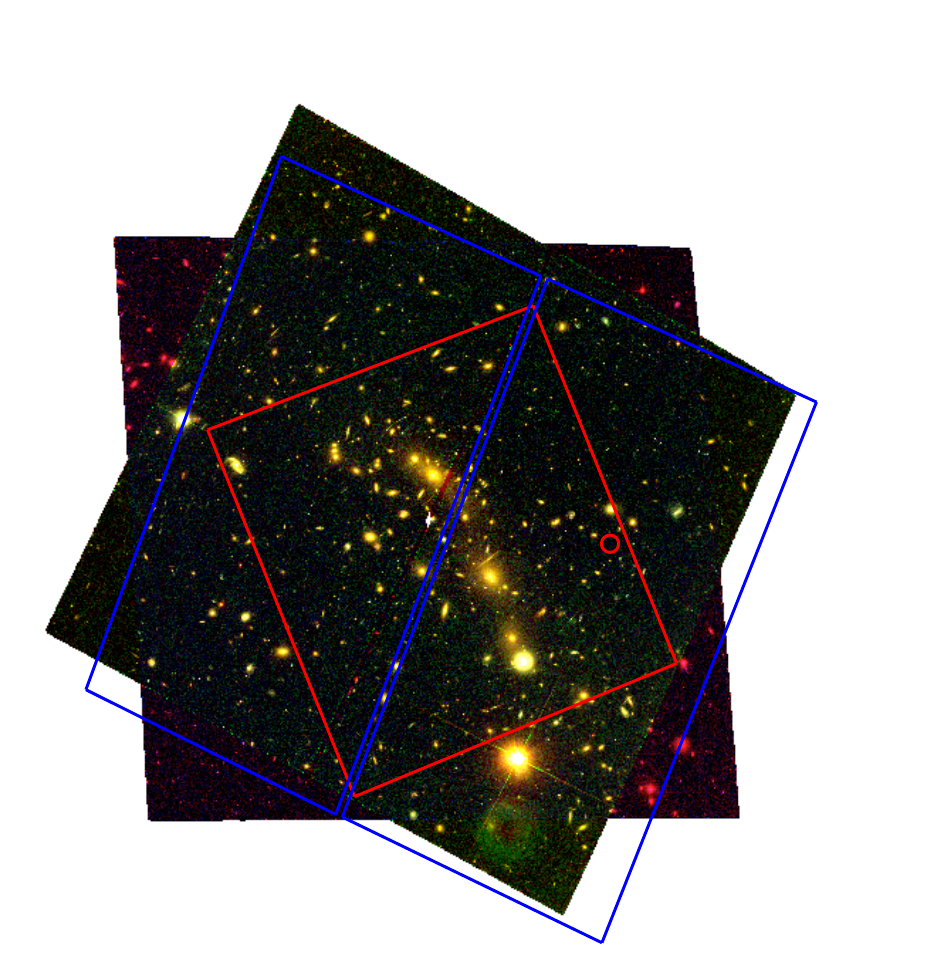}
\includegraphics[width=0.33\textwidth,clip,trim=60mm 100mm 60mm 80mm]{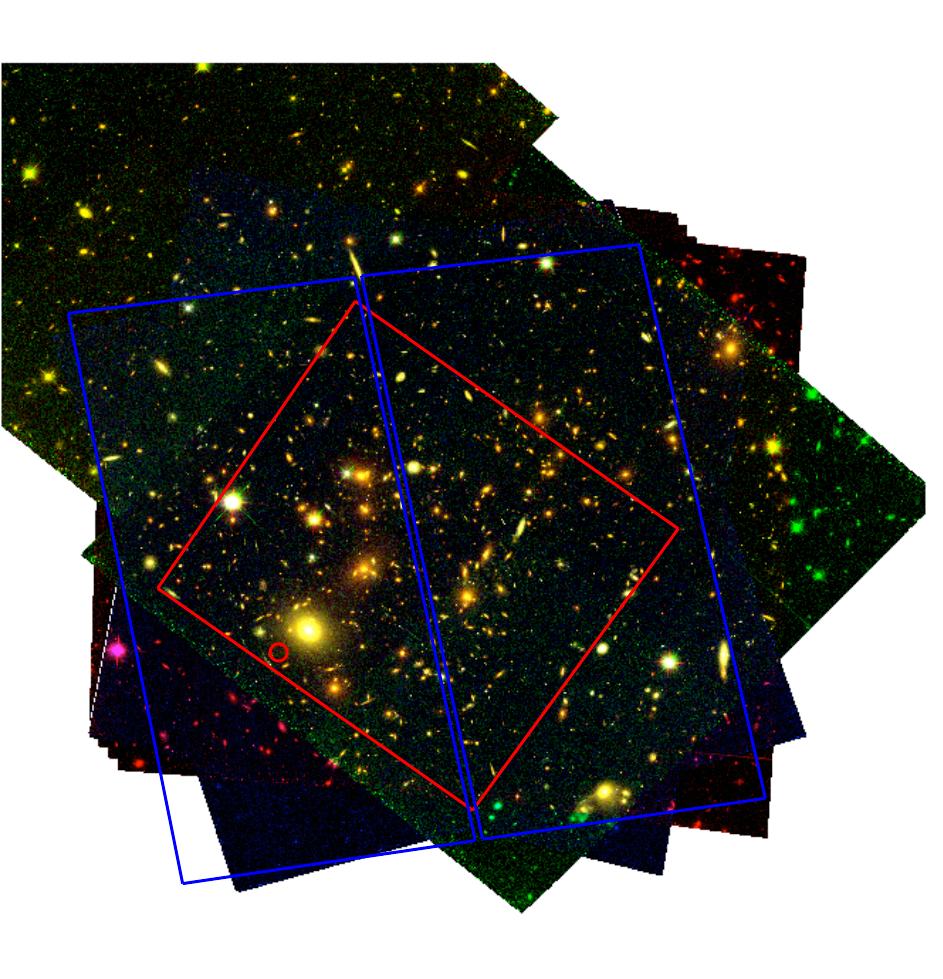}\\
\includegraphics[width=0.33\textwidth,clip,trim=60mm 100mm 60mm 80mm]{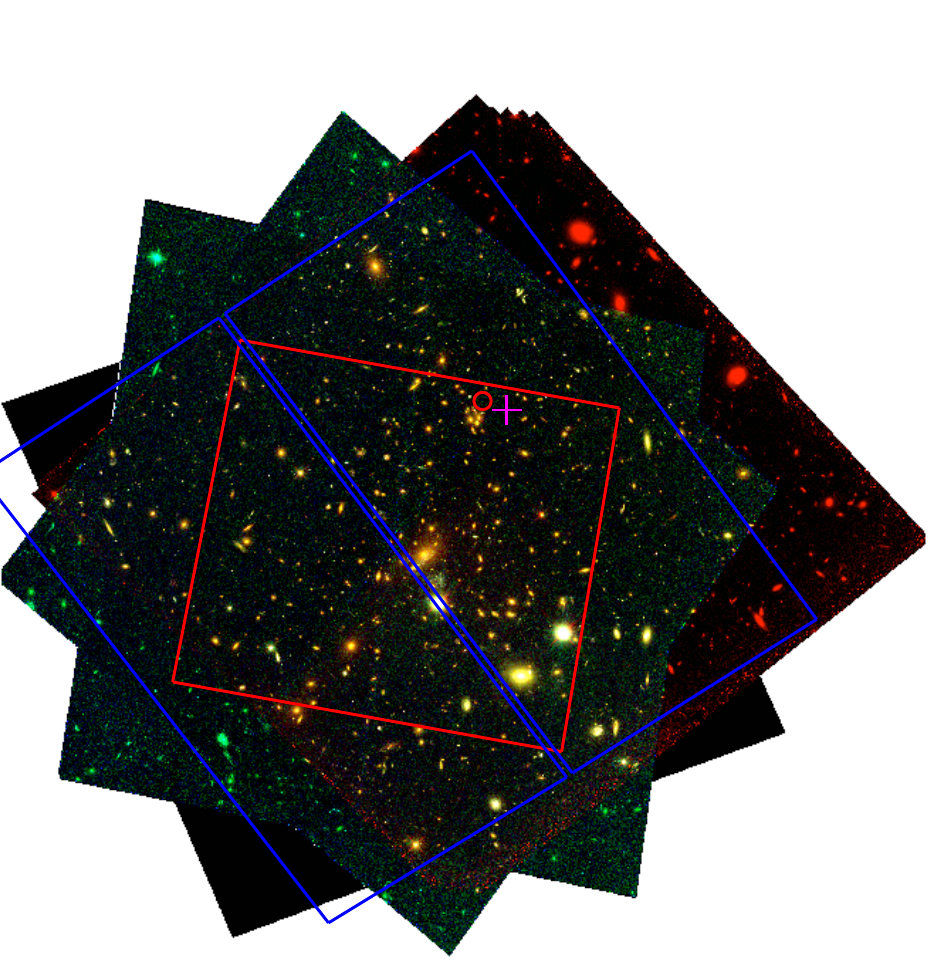}
\includegraphics[width=0.33\textwidth,clip,trim=60mm 100mm 60mm 80mm]{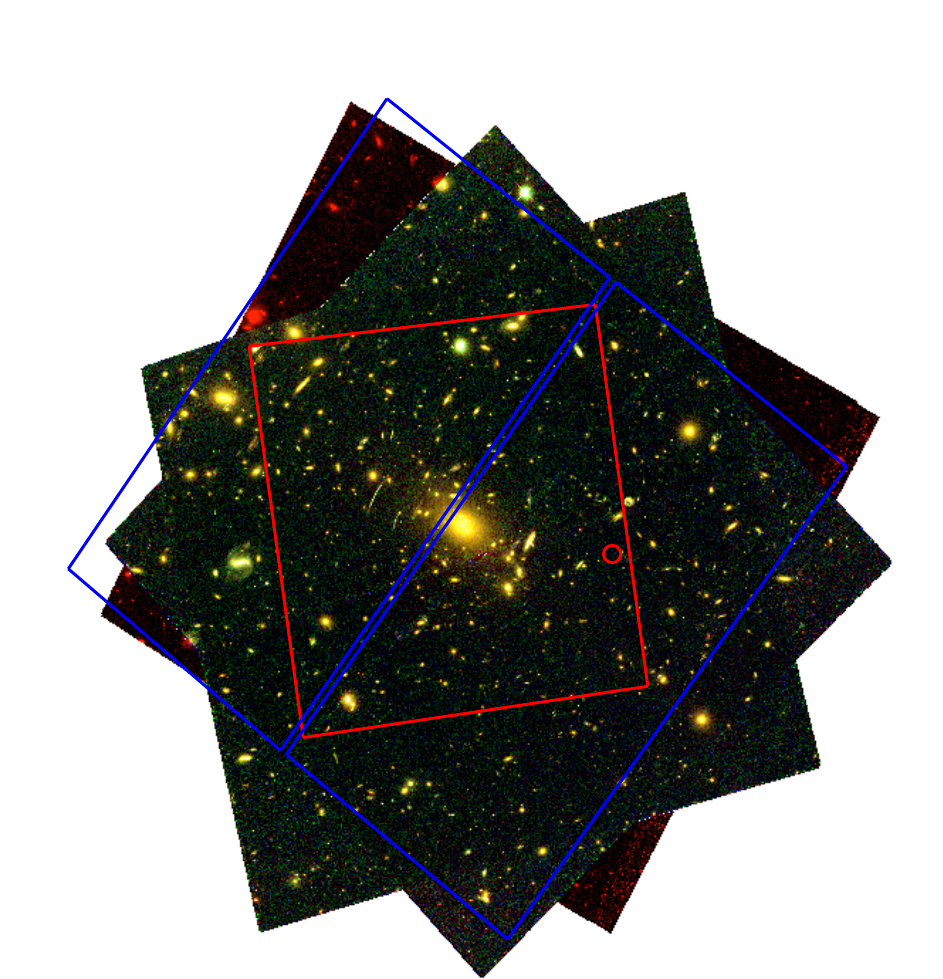}
\includegraphics[width=0.33\textwidth,clip,trim=45mm 40mm 60mm 0mm]{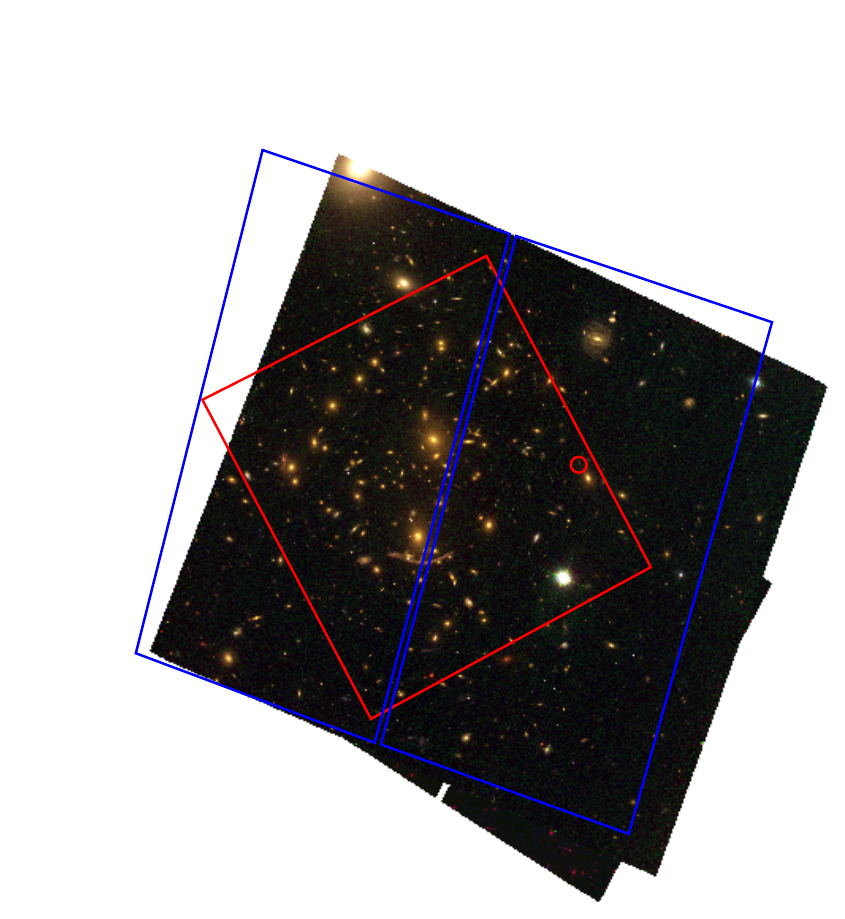}
\caption{Color images (F435W+F606W+F814W) of the archival \textit{HST} data for the HFF (except for A370 where we show F475W+F625W+F814W). Overlaid in blue (red) are the apertures of the planned, deep HFF observations with ACS (WFC3). In all cases the parallel flanking fields -- which fall well outside the depicted area -- have not been previously imaged with \textit{HST}. The images shown are trimmed to the regions in which data in at least two filters are available whose central wavelengths differ by at least 150\AA, thus providing color information suitable for the identification of multiple-image systems. \label{fig:hff-layout}} 
\end{figure*}

Fig.~\ref{fig:hff-layout} shows the outlines of the area targeted by the planned HFF observations overlaid on the existing \textit{HST} imaging data. Note that, in all cases, the blank flanking fields fall outside the area covered by the existing data; hence, the gravitational magnification induced at the location of the flanking fields by either the clusters themselves, or by large-scale structure in their immediate vicinity, can presently not be constrained with \textit{HST} data.

\section{Strong-lensing features}
\subsection{Identifications}
\label{sec-multiples}
We now review, for each cluster, the associations of multiple images found in these HST images prior to HFF observations and used to constrain the strong lensing models (see Section \ref{sec-sl}). 
We adopt the term `system'  for a set of multiple images arising from the same source, and the notation X.Y to describe image Y of system X.
We build our identification of multiple systems from earlier published lists of multiple images (identified either by our own or by other groups). For clusters having 
such a large number of constraints our parametric analysis is quite sensitive to wrong identifications: we therefore added iteratively new systems and used the current 
model to predict additional counter-images. In a few cases where the identification was ambiguous between multiple candidates, these counter-images were not added as constraints. 

\subsubsection{Abell 2744}

\begin{figure*}
\includegraphics[width=\textwidth]{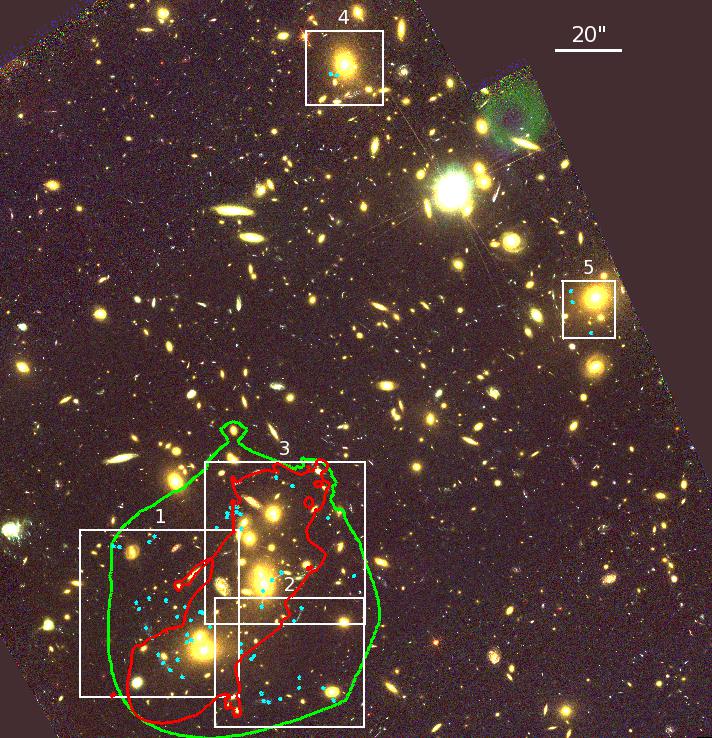}
\caption{\label{fig:a2744-imlarge}
Color image of A2744 as obtained with HST/ACS (F435W+F606W+F814W). North is up and East is left. Shown in red is the critical line at $z=7$, in green the enclosing region where we expect multiple images at $z=7$, and blue circles mark the location of multiple images used as constraints in the modelling. Thin white lines delineate the regions shown in more detail in Fig.~\ref{fig:a2744-imzoom}.}
\end{figure*}

\begin{figure*}
\includegraphics[width=0.49\textwidth]{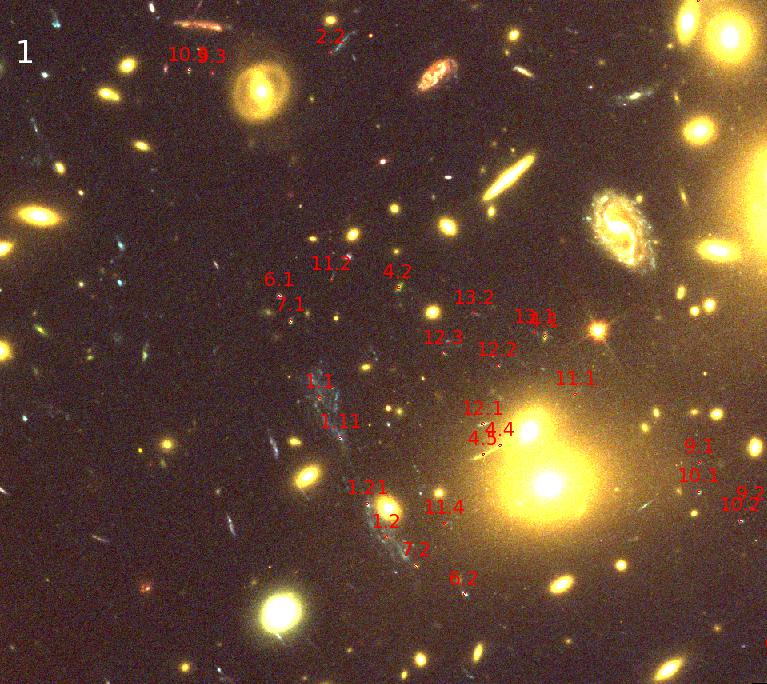}
\includegraphics[width=0.49\textwidth]{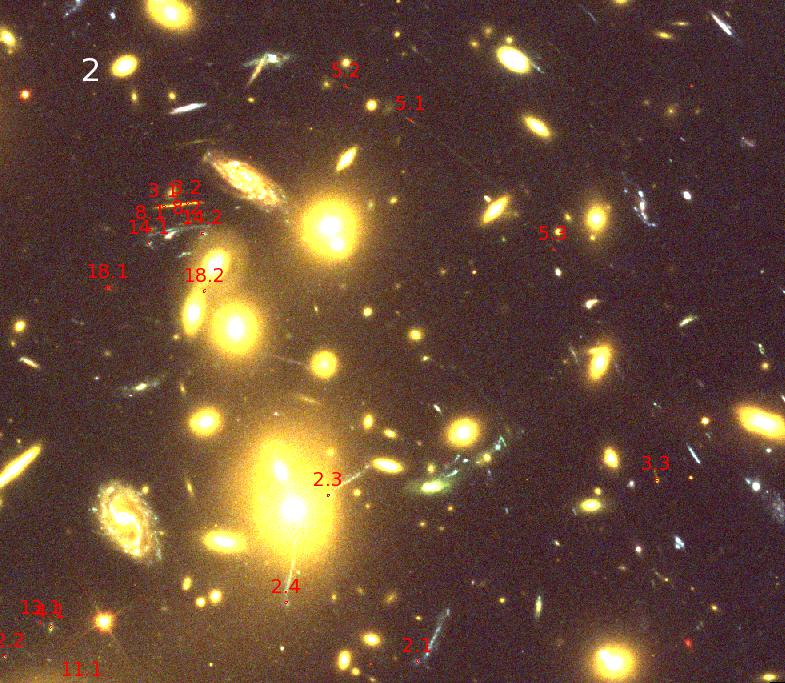}\\[1mm]
\includegraphics[width=0.49\textwidth]{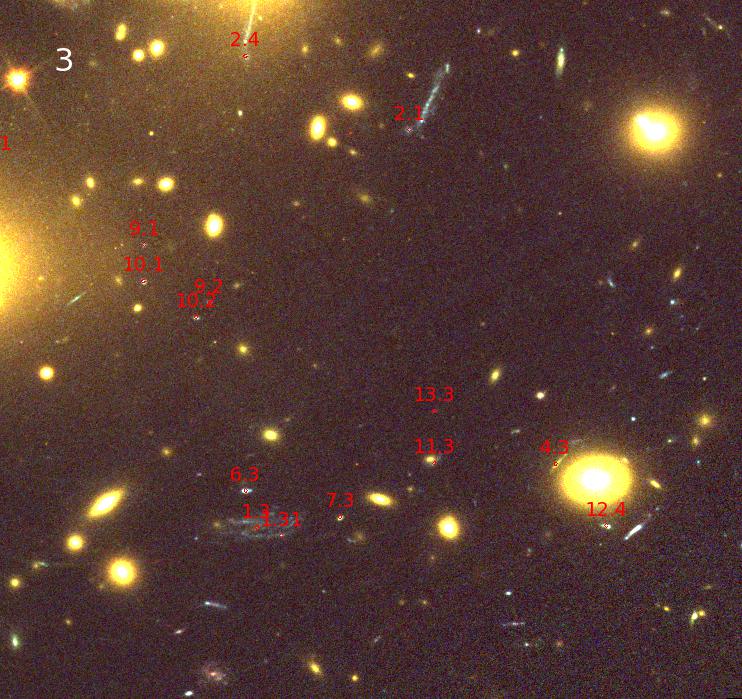}
\includegraphics[width=0.29\textwidth]{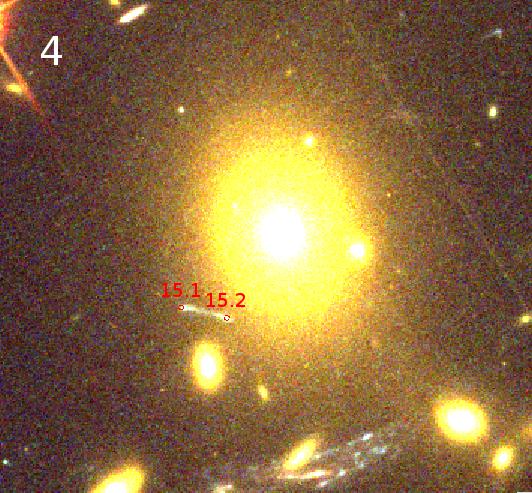}
\includegraphics[width=0.19\textwidth]{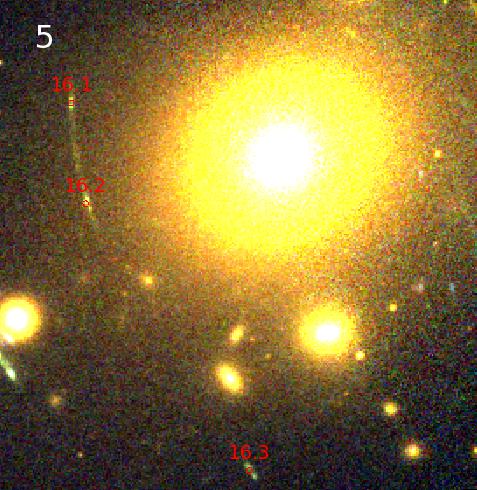}
\caption{\label{fig:a2744-imzoom}
Sub-regions of A2744 as marked and labeled in Fig.~\ref{fig:a2744-imlarge}.  Strong-lensing features as listed in Table~\ref{tab:a2744-SL} are labeled.}
\end{figure*}

An earlier strong lensing analysis of this cluster was previously published by \citet{Merten}, who identified 11 systems and a total of 34 images. We build 
from these 11 systems and identify 7 additional, convincing systems of multiple images (Fig. \ref{fig:a2744-imzoom}), leading to a total of 55 images. The majority of these 
new systems are located around the core of the cluster, but 3 new systems are identified around several sub-clumps (galaxy-scale) groups close to the North and 
North-West limits  of the ACS coverage (Fig. \ref{fig:a2744-imlarge}). These sub-clumps were identified as 'N' and 'NW', respectively, in \citet{Merten}. 

Compared to \citet{Merten} we removed image 8.3 from our list of constraints, as its colours and location do not match well for a multiple system (in particular, the 
geometry should be similar to system 3). In addition, we do not find any convincing counter-image (third image of a triple 'fold' configuration) for systems 14, 15 
and 18.

\begin{table}
\begin{tabular}{lllll}
ID & $\alpha$ & $\delta$ & $z_{\rm prior}$ & $z_{\rm model}$ \\
   & [deg] & [deg] & & \\
\hline 
 1.1 & 00:14:23.41 & -30:24:14.10 & [0.6-6.0] & 1.72$\pm$0.07\\
 1.2 & 00:14:23.03 & -30:24:24.56 &  & \\
 1.3 & 00:14:20.69 & -30:24:35.95 &  & \\
 2.1 & 00:14:19.98 & -30:24:12.06 & [0.6-6.0] & 2.33$\pm$0.12\\
 2.2 & 00:14:23.35 & -30:23:48.21 &  & \\
 2.3 & 00:14:20.50 & -30:23:59.63 &  & \\
 2.4 & 00:14:20.74 & -30:24:07.66 &  & \\
 3.1 & 00:14:21.45 & -30:23:37.95 & [0.6-6.0] & 2.34$\pm$0.13\\
 3.2 & 00:14:21.31 & -30:23:37.69 &  & \\
 3.3 & 00:14:18.60 & -30:23:58.44 &  & \\
 4.1 & 00:14:22.11 & -30:24:09.48 & 3.58 & \\
 4.2 & 00:14:22.95 & -30:24:05.84 & 3.58 & \\
 4.3 & 00:14:19.30 & -30:24:32.13 & 3.58 & \\
 4.4 & 00:14:22.37 & -30:24:17.69 & 3.58 & \\
 4.5 & 00:14:22.46 & -30:24:18.38 & 3.58 & \\
 5.1 & 00:14:20.02 & -30:23:31.45 & [0.6-6.0] & 3.50$\pm$0.47\\
 5.2 & 00:14:20.40 & -30:23:28.95 &  & \\
 5.3 & 00:14:19.19 & -30:23:41.14 &  & \\
 6.1 & 00:14:23.65 & -30:24:06.48 & 2.019 & \\
 6.2 & 00:14:22.57 & -30:24:28.84 & 2.019 & \\
 6.3 & 00:14:20.74 & -30:24:33.74 & 2.019 & \\
 7.1 & 00:14:23.58 & -30:24:08.35 & [0.6-6.0] & 2.63$\pm$0.17\\
 7.2 & 00:14:22.85 & -30:24:26.73 &  & \\
 7.3 & 00:14:20.30 & -30:24:35.33 &  & \\
 8.1 & 00:14:21.53 & -30:23:39.62 & [0.6-6.0] & 5.12$\pm$0.79\\
 8.2 & 00:14:21.32 & -30:23:39.20 &  & \\
 9.1 & 00:14:21.21 & -30:24:18.98 & [0.6-6.0] & 4.60$\pm$0.35\\
 9.2 & 00:14:20.91 & -30:24:22.47 &  & \\
 9.3 & 00:14:24.04 & -30:23:49.75 &  & \\
 10.1 & 00:14:21.22 & -30:24:21.16 & [0.6-6.0] & 6.00$\pm$0.25\\
 10.2 & 00:14:20.97 & -30:24:23.33 &  & \\
 10.3 & 00:14:24.17 & -30:23:49.56 &  & \\
 11.1 & 00:14:21.93 & -30:24:13.89 & [0.6-6.0] & 2.88$\pm$0.16\\
 11.2 & 00:14:23.34 & -30:24:05.23 &  & \\
 11.3 & 00:14:19.87 & -30:24:32.09 &  & \\
 11.4 & 00:14:22.69 & -30:24:23.55 &  & \\
 12.1 & 00:14:22.47 & -30:24:16.09 & [0.6-6.0] & 4.77$\pm$0.32\\
 12.2 & 00:14:22.38 & -30:24:11.72 &  & \\
 12.3 & 00:14:22.70 & -30:24:10.76 &  & \\
 12.4 & 00:14:19.07 & -30:24:35.83 &  & \\
 13.1 & 00:14:22.17 & -30:24:09.21 & [0.6-6.0] & 1.51$\pm$0.05\\
 13.2 & 00:14:22.51 & -30:24:07.79 &  & \\
 13.3 & 00:14:19.87 & -30:24:28.96 &  & \\
 14.1 & 00:14:21.54 & -30:23:40.69 & [0.6-6.0] & 3.86$\pm$0.84\\
 14.2 & 00:14:21.23 & -30:23:39.97 &  & \\
 15.1 & 00:14:19.14 & -30:21:27.83 & [0.6-6.0] & 5.82$\pm$0.69\\
 15.2 & 00:14:18.99 & -30:21:28.25 &  & \\
 16.1 & 00:14:13.57 & -30:22:32.91 & [0.6-6.0] & 4.70$\pm$0.58\\
 16.2 & 00:14:13.53 & -30:22:36.36 &  & \\
 16.3 & 00:14:13.10 & -30:22:45.51 &  & \\
 18.1 & 00:14:21.78 & -30:23:44.02 & [0.6-6.0] & 3.37$\pm$0.69\\
 18.2 & 00:14:21.21 & -30:23:44.29 &  & \\
\hline
\end{tabular}

\caption{\label{tab:a2744-SL}Strong-lensing features identified in the existing \textit{HST} images of A\,2744. Spectroscopic redshifts are listed under $z_{\rm prior}$ where available; other constraints used during the optimization process are quoted as a range, enclosed in square brackets.}
\end{table}

\subsubsection{MACS\,J0416}

\begin{figure*}
\includegraphics[width=\textwidth]{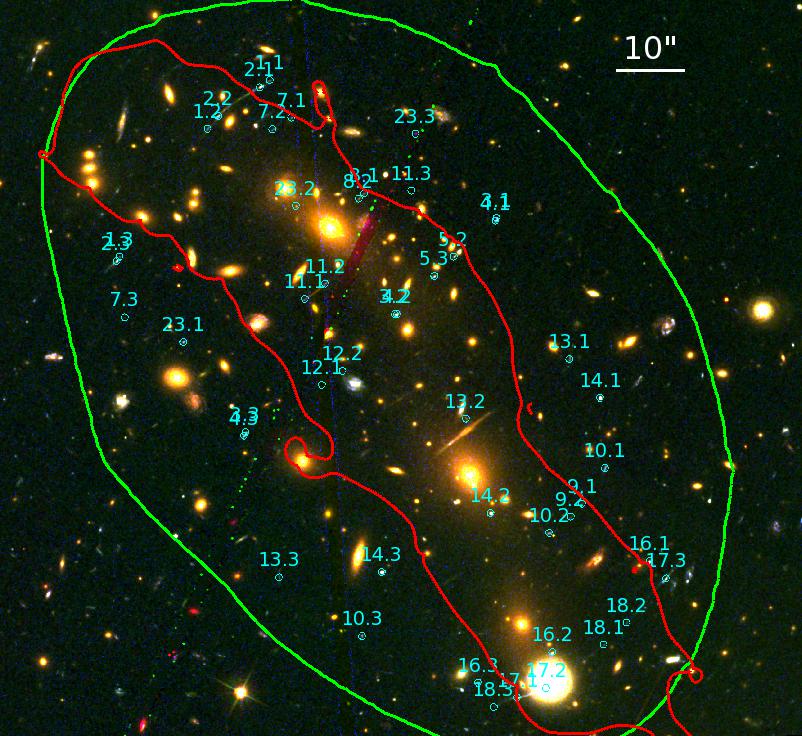}
\caption{\label{fig:m0416-imlarge} Same as Fig.~\ref{fig:a2744-imlarge} but for MACS\,J0416 and filter combination  F435W+F606W+F814W.}
\end{figure*}

\citet{Zitrin2013} presented a strong-lensing model with 23 systems identified in this cluster in the new CLASH \citep{Postman2012} data, producing in total 70 images. 
Due to the very elongated nature of the cluster mass distribution (Fig. \ref{fig:m0416-imlarge}), all systems appear as triply imaged configurations. Among the total list of 
23 systems, 10 systems and 36 images were considered by \citet{Zitrin2013} as less robust candidate multiple images. 

We have built from this list of strong-lensing features and selected only the most robust systems and a few candidate systems showing clear counterimages based 
on our preliminary strong lensing analysis (Table \ref{tab:m0416-SL}). New spectroscopic information (see Sect. \ref{spec-m0416}) was also incorporated to help with this selection.
In total, our final list contains 17 systems and 47 images. We do not identify a reliable (unambiguous)
3rd image for systems 5, 8, 9 and 12.

\begin{table}
\begin{tabular}{lllll}
ID & $\alpha$ & $\delta$ & $z_{\rm prior}$ & $z_{\rm model}$ \\
   & [deg] & [deg] & & \\
\hline 
 1.1 & 04:16:09.78 & -24:03:41.73 & 1.896 & \\
 1.2 & 04:16:10.43 & -24:03:48.75 & 1.896 & \\
 1.3 & 04:16:11.36 & -24:04:07.21 & 1.896 & \\
 2.1 & 04:16:09.88 & -24:03:42.77 & 1.8925 & \\
 2.2 & 04:16:10.32 & -24:03:46.93 & 1.8925 & \\
 2.3 & 04:16:11.39 & -24:04:07.86 & 1.8925 & \\
 3.1 & 04:16:07.39 & -24:04:01.62 & 1.9885 & \\
 3.2 & 04:16:08.46 & -24:04:15.53 & 1.9885 & \\
 3.3 & 04:16:10.04 & -24:04:32.56 & 1.9885 & \\
 4.1 & 04:16:07.40 & -24:04:02.01 & [0.6-6.0] & 2.04$\pm$0.08\\
 4.2 & 04:16:08.44 & -24:04:15.53 &  & \\
 4.3 & 04:16:10.05 & -24:04:33.08 &  & \\
 5.2 & 04:16:07.84 & -24:04:07.21 & [0.6-6.0] & 1.72$\pm$0.20\\
 5.3 & 04:16:08.04 & -24:04:10.01 &  & \\
 7.1 & 04:16:09.55 & -24:03:47.13 & 2.0854 & \\
 7.2 & 04:16:09.75 & -24:03:48.82 & 2.0854 & \\
 7.3 & 04:16:11.31 & -24:04:15.99 & 2.0854 & \\
 8.1 & 04:16:08.78 & -24:03:58.05 & [0.6-6.0] & 3.19$\pm$1.96\\
 8.2 & 04:16:08.84 & -24:03:58.83 &  & \\
 9.1 & 04:16:06.49 & -24:04:42.90 & [0.6-6.0] & 2.73$\pm$0.47\\
 9.2 & 04:16:06.61 & -24:04:44.78 &  & \\
 10.1 & 04:16:06.24 & -24:04:37.76 & 2.2982 & \\
 10.2 & 04:16:06.83 & -24:04:47.12 & 2.2982 & \\
 10.3 & 04:16:08.81 & -24:05:02.04 & 2.2982 & \\
 11.1 & 04:16:09.41 & -24:04:13.32 & [0.6-6.0] & 1.08$\pm$0.04\\
 11.2 & 04:16:09.20 & -24:04:11.11 &  & \\
 11.3 & 04:16:08.29 & -24:03:57.69 &  & \\
 12.1 & 04:16:09.23 & -24:04:25.74 & [0.6-6.0] & 1.63$\pm$0.24\\
 12.2 & 04:16:09.01 & -24:04:23.72 &  & \\
 13.1 & 04:16:06.62 & -24:04:22.03 & 3.2226 & \\
 13.2 & 04:16:07.71 & -24:04:30.61 & 3.2226 & \\
 13.3 & 04:16:09.68 & -24:04:53.56 & 3.2226 & \\
 14.1 & 04:16:06.30 & -24:04:27.62 & 2.0531 & \\
 14.2 & 04:16:07.45 & -24:04:44.26 & 2.0531 & \\
 14.3 & 04:16:08.60 & -24:04:52.78 & 2.0531 & \\
 16.1 & 04:16:05.77 & -24:04:51.22 & [0.6-6.0] & 2.09$\pm$0.08\\
 16.2 & 04:16:06.80 & -24:05:04.35 &  & \\
 16.3 & 04:16:07.58 & -24:05:08.77 &  & \\
 17.1 & 04:16:07.17 & -24:05:10.91 & 2.2181 & \\
 17.2 & 04:16:06.87 & -24:05:09.55 & 2.2181 & \\
 17.3 & 04:16:05.60 & -24:04:53.69 & 2.2181 & \\
 18.1 & 04:16:06.26 & -24:05:03.24 & [0.6-6.0] & 2.19$\pm$0.10\\
 18.2 & 04:16:06.02 & -24:05:00.06 &  & \\
 18.3 & 04:16:07.42 & -24:05:12.28 &  & \\
 23.1 & 04:16:10.69 & -24:04:19.56 & [0.6-6.0] & 2.25$\pm$0.11\\
 23.2 & 04:16:09.50 & -24:03:59.87 &  & \\
 23.3 & 04:16:08.24 & -24:03:49.47 &  & \\
\hline
\end{tabular}

\caption{\label{tab:m0416-SL}
As Table~\ref{tab:a2744-SL} but for MACS\,J0416.}
\end{table}

\subsubsection{MACS\,J0717}

The set of multiply imaged systems used in the analysis of MACS\,J0717
mainly follows the one described in \citet{2012A&A...544A..71L}, with a few exceptions
that we discuss here.
The location of image 1.5 has been revisited, following \citet{Medezinski2013}.
System 2 has been removed from the analysis, after discussion between the different
teams, given its faintness.

In addition, we added new spectroscopic redshifts measurements, in agreement with 
our previous estimates, obtained by the GLASS survey \citep{2014ApJ...782L..36S} for 
three multiple systems (4,6 and 12).
Finally, \citet{2014ApJ...783L..12V} recently measured a spectroscopic redshift 
$z=6.4$ for two galaxies which we confirm to be multiply imaged with 3 images 
identified based on the  mass model. We add this new constraint as system 19.

\begin{table}
\begin{tabular}{lllll}
ID & $\alpha$ & $\delta$ & $z_{\rm prior}$ & $z_{\rm model}$ \\
   & [deg] & [deg] & & \\
\hline 
 1.1 & 07:17:34.86 & 37:44:28.39 & 2.963 & \\
 1.2 & 07:17:34.51 & 37:44:24.43 & 2.963 & \\
 1.3 & 07:17:33.82 & 37:44:17.91 & 2.963 & \\
 1.4 & 07:17:32.23 & 37:44:13.14 & 2.963 & \\
 1.5 & 07:17:37.39 & 37:45:40.90 & 2.963 & \\
 2.1 & 07:17:34.26 & 37:44:27.78 & [0.6-6.0] & 3.14$\pm$1.12\\
 2.2 & 07:17:33.69 & 37:44:21.30 &  & \\
 3.1 & 07:17:35.64 & 37:44:29.44 & 1.855 & \\
 3.2 & 07:17:34.66 & 37:44:21.11 & 1.855 & \\
 3.3 & 07:17:37.70 & 37:45:13.86 & 1.855 & \\
 4.1 & 07:17:31.44 & 37:45:01.57 & 1.855 & \\
 4.2 & 07:17:30.32 & 37:44:40.72 & 1.855 & \\
 4.3 & 07:17:33.83 & 37:45:47.80 & 1.855 & \\
 5.2 & 07:17:30.69 & 37:44:34.19 & [0.6-6.0] & 4.28$\pm$0.27\\
 5.1 & 07:17:31.17 & 37:44:48.74 &  & \\
 5.3 & 07:17:36.00 & 37:46:02.77 &  & \\
 6.1 & 07:17:27.43 & 37:45:25.59 & 2.393 & \\
 6.2 & 07:17:27.04 & 37:45:09.93 & 2.393 & \\
 6.3 & 07:17:29.73 & 37:46:11.21 & 2.393 & \\
 7.1 & 07:17:27.97 & 37:45:58.90 & [0.6-6.0] & 1.83$\pm$0.56\\
 7.2 & 07:17:27.61 & 37:45:50.88 &  & \\
 8.1 & 07:17:27.98 & 37:46:10.83 & [0.6-6.0] & 2.98$\pm$0.19\\
 8.2 & 07:17:26.89 & 37:45:47.43 &  & \\
 8.3 & 07:17:25.55 & 37:45:06.70 &  & \\
 12.1 & 07:17:32.44 & 37:45:06.82 & 1.699 & \\
 12.2 & 07:17:30.62 & 37:44:34.52 & 1.699 & \\
 12.3 & 07:17:33.89 & 37:45:38.39 & 1.699 & \\
 13.1 & 07:17:32.52 & 37:45:02.32 & 2.547 & \\
 13.2 & 07:17:30.61 & 37:44:22.86 & 2.547 & \\
 13.3 & 07:17:35.08 & 37:45:48.21 & 2.547 & \\
 14.1 & 07:17:33.30 & 37:45:07.96 & 1.855 & \\
 14.2 & 07:17:31.11 & 37:44:22.92 & 1.855 & \\
 14.3 & 07:17:35.08 & 37:45:37.21 & 1.855 & \\
 15.1 & 07:17:28.25 & 37:46:19.26 & 2.405 & \\
 15.2 & 07:17:26.09 & 37:45:36.32 & 2.405 & \\
 15.3 & 07:17:25.58 & 37:45:16.20 & 2.405 & \\
 16.1 & 07:17:28.59 & 37:46:23.89 & [0.6-6.0] & 3.71$\pm$0.30\\
 16.2 & 07:17:26.05 & 37:45:34.51 &  & \\
 16.3 & 07:17:25.66 & 37:45:13.44 &  & \\
 17.3 & 07:17:25.97 & 37:45:12.74 & [0.6-6.0] & 2.79$\pm$0.22\\
 17.2 & 07:17:26.26 & 37:45:31.82 &  & \\
 17.1 & 07:17:28.65 & 37:46:18.58 &  & \\
 18.1 & 07:17:27.41 & 37:46:07.14 & [0.6-6.0] & 1.88$\pm$0.60\\
 18.2 & 07:17:26.68 & 37:45:51.69 &  & \\
 19.1 & 07:17:38.17 & 37:45:16.87 & 6.40 & \\
 19.2 & 07:17:37.86 & 37:44:33.87 & 6.40 & \\
 19.3 & 07:17:31.45 & 37:43:53.78 & 6.40 & \\
\hline
\end{tabular}

\caption{\label{tab:m0717-SL}
As Table~\ref{tab:a2744-SL} but for MACS\,J0717.}
\end{table}

\begin{figure*}
\includegraphics[width=\textwidth]{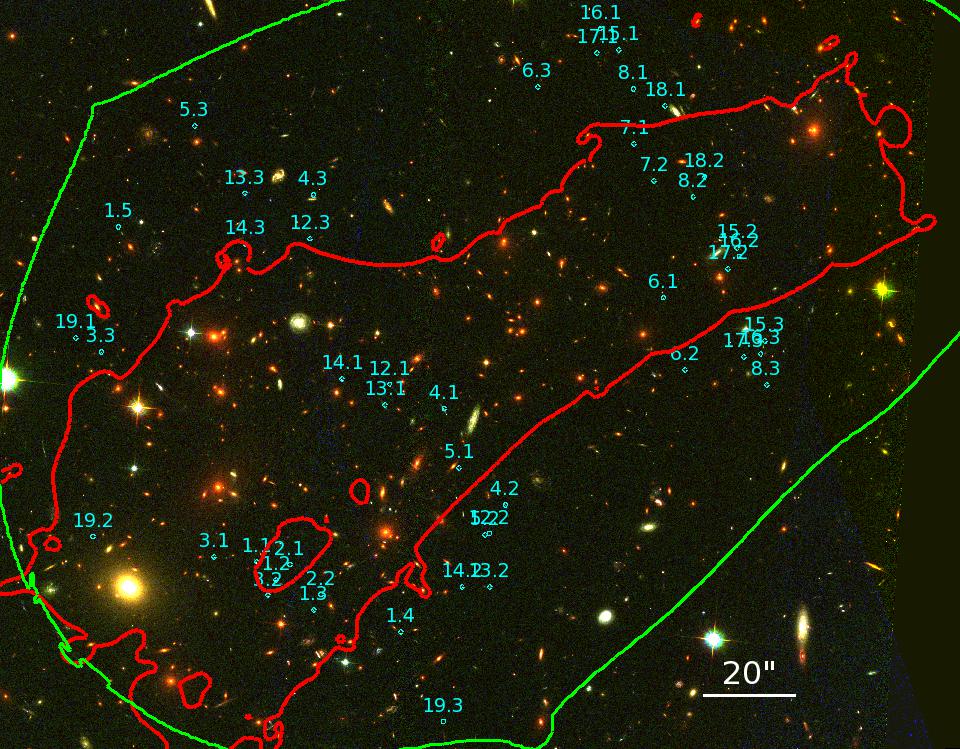}
\caption{\label{fig:m0717-imlarge}
Same as Fig.~\ref{fig:a2744-imlarge} but for MACS\,J0717 and filter combination  F435W+F555W+F814W.}
\end{figure*}

\subsubsection{MACS\,J1149}

The set of multiply imaged systems used in the analysis of MACS\,J1149
is based on the one presented by \citet{Smith09}, with the addition of
six new systems as proposed by \citet{Zitrin2011}: systems 5, 6, 7, 8, 13 and 14.

\begin{table}
\begin{tabular}{lllll}
ID & $\alpha$ & $\delta$ & $z_{\rm prior}$ & $z_{\rm model}$ \\
   & [deg] & [deg] & & \\
\hline 
 1.1 & 11:49:35.28 & 22:23:45.63 & 1.480 & \\
 1.2 & 11:49:35.86 & 22:23:50.78 & 1.480 & \\
 1.3 & 11:49:36.82 & 22:24:08.73 & 1.480 & \\
 2.1 & 11:49:36.58 & 22:23:23.06 & 1.894 & \\
 2.2 & 11:49:37.46 & 22:23:32.94 & 1.894 & \\
 2.3 & 11:49:37.58 & 22:23:34.37 & 1.894 & \\
 3.1 & 11:49:33.81 & 22:23:59.60 & 2.497 & \\
 3.2 & 11:49:34.25 & 22:24:11.07 & 2.497 & \\
 3.3 & 11:49:36.31 & 22:24:25.85 & 2.497 & \\
 4.1 & 11:49:34.32 & 22:23:48.57 & [0.6-6.0] & 2.57$\pm$0.15\\
 4.2 & 11:49:34.66 & 22:24:02.62 &  & \\
 4.3 & 11:49:37.01 & 22:24:22.03 &  & \\
 5.1 & 11:49:35.94 & 22:23:35.02 & [0.6-6.0] & 2.61$\pm$0.29\\
 5.2 & 11:49:36.27 & 22:23:37.77 &  & \\
 5.3 & 11:49:37.91 & 22:24:12.74 &  & \\
 6.1 & 11:49:35.93 & 22:23:33.16 & [0.6-6.0] & 2.59$\pm$0.27\\
 6.2 & 11:49:36.43 & 22:23:37.89 &  & \\
 6.3 & 11:49:37.93 & 22:24:09.02 &  & \\
 7.1 & 11:49:35.75 & 22:23:28.80 & [0.6-6.0] & 2.54$\pm$0.25\\
 7.2 & 11:49:36.81 & 22:23:39.37 &  & \\
 7.3 & 11:49:37.82 & 22:24:04.47 &  & \\
 8.1 & 11:49:35.64 & 22:23:39.66 & [0.6-6.0] & 3.10$\pm$0.41\\
 8.2 & 11:49:35.95 & 22:23:42.20 &  & \\
 8.3 & 11:49:37.69 & 22:24:19.99 &  & \\
 9.1 & 11:49:37.24 & 22:25:34.40 & [0.6-6.0] & 4.03$\pm$0.82\\
 9.2 & 11:49:36.93 & 22:25:37.98 &  & \\
 9.3 & 11:49:36.78 & 22:25:38.00 &  & \\
 10.1 & 11:49:37.07 & 22:25:31.83 & [0.6-6.0] & 4.13$\pm$0.66\\
 10.2 & 11:49:36.87 & 22:25:32.26 &  & \\
 10.3 & 11:49:36.53 & 22:25:35.80 &  & \\
 13.1 & 11:49:36.89 & 22:23:52.03 & [0.6-6.0] & 1.31$\pm$0.04\\
 13.2 & 11:49:36.68 & 22:23:47.96 &  & \\
 13.3 & 11:49:36.01 & 22:23:37.89 &  & \\
 14.1 & 11:49:34.00 & 22:24:12.61 & [0.6-6.0] & 2.85$\pm$0.58\\
 14.2 & 11:49:33.80 & 22:24:09.45 &  & \\
\hline
\end{tabular}

\caption{\label{tab:m1149-SL}As Table~\ref{tab:a2744-SL} but for MACS\,J1149.}
\end{table}

\begin{figure*}
\includegraphics[width=\textwidth]{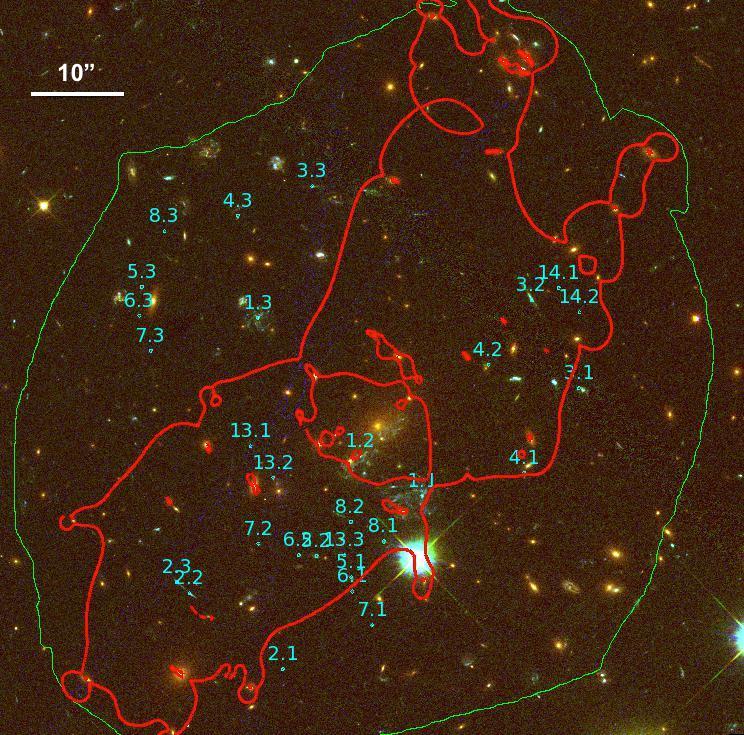}
\caption{\label{fig:m1149-imlarge}
Same as Fig.~\ref{fig:a2744-imlarge} but for MACS\,J1149 and filter combination  F435W+F555W+F814W.}
\end{figure*}

\subsubsection{Abell S1063}

We identify in this cluster  14 multiply imaged systems, all in a configuration of 3 images 
 except for System 6 which shows as a quad \citep{Boone2013}. The robustness 
of these systems has been checked with a dedicated spectroscopic follow-up (Sect. \ref{spec-as1063}). 
 While this 
paper was being written, \citet{Monna} presented their own identifications, partially overlapping
 with our own list: we identify 4 additional systems and they identified 2 additional candidate 
 systems which appear much fainter and less robust. In total, our list comprises 41 images of the 14 systems, 
 as we do not find an unambiguous counterpart for systems 10 and 12.
 
\begin{table}\label{tab:r2248-SL}
\begin{tabular}{lllll}
ID & $\alpha$ & $\delta$ & $z_{\rm prior}$ & $z_{\rm model}$ \\
   & [deg] & [deg] & & \\
\hline 
 1.1 & 22:48:46.68 & -44:31:37.19 & 1.235 & \\
 1.2 & 22:48:47.02 & -44:31:44.27 & 1.235 & \\
 1.3 & 22:48:44.75 & -44:31:16.31 & 1.235 & \\
 2.1 & 22:48:41.81 & -44:31:41.96 & 1.429 & \\
 2.2 & 22:48:42.21 & -44:31:57.19 & 1.429 & \\
 2.3 & 22:48:45.24 & -44:32:23.93 & 1.429 & \\
 3.1 & 22:48:45.08 & -44:31:38.33 & 1.398 & \\
 3.2 & 22:48:43.01 & -44:31:24.93 & 1.398 & \\
 3.3 & 22:48:46.36 & -44:32:11.54 & 1.398 & \\
 4.1 & 22:48:46.25 & -44:31:52.09 & 1.260 & \\
 4.2 & 22:48:46.13 & -44:31:47.78 & 1.260 & \\
 4.3 & 22:48:43.17 & -44:31:17.64 & 1.260 & \\
 5.1 & 22:48:42.02 & -44:32:27.69 & [0.6-6.0] & 2.30$\pm$0.10\\
 5.2 & 22:48:41.56 & -44:32:23.94 &  & \\
 5.3 & 22:48:39.74 & -44:31:46.31 &  & \\
 6.1 & 22:48:45.37 & -44:31:48.06 & 6.107 & \\
 6.2 & 22:48:45.81 & -44:32:14.86 & 6.107 & \\
 6.3 & 22:48:43.45 & -44:32:04.66 & 6.107 & \\
 6.4 & 22:48:41.11 & -44:31:11.41 & 6.107 & \\
 7.1 & 22:48:42.92 & -44:32:09.16 & [0.6-6.0] & 3.10$\pm$0.14\\
 7.2 & 22:48:44.98 & -44:32:19.32 &  & \\
 7.3 & 22:48:40.96 & -44:31:19.55 &  & \\
 8.1 & 22:48:46.01 & -44:31:49.92 & [0.6-6.0] & 2.33$\pm$0.08\\
 8.2 & 22:48:46.21 & -44:32:03.94 &  & \\
 8.3 & 22:48:42.22 & -44:31:10.75 &  & \\
 9.1 & 22:48:43.28 & -44:32:27.02 & [0.6-6.0] & 2.34$\pm$0.10\\
 9.2 & 22:48:41.94 & -44:32:18.91 &  & \\
 9.3 & 22:48:40.27 & -44:31:34.63 &  & \\
 10.1 & 22:48:39.90 & -44:32:01.16 & [0.6-6.0] & 2.30$\pm$0.13\\
 10.3 & 22:48:42.68 & -44:32:35.07 &  & \\
 11.1 & 22:48:44.60 & -44:32:19.90 & [0.6-6.0] & 3.56$\pm$0.19\\
 11.2 & 22:48:42.92 & -44:32:12.25 &  & \\
 11.3 & 22:48:40.75 & -44:31:19.12 &  & \\
 12.1 & 22:48:41.32 & -44:32:11.83 & [0.6-6.0] & 3.46$\pm$0.40\\
 12.2 & 22:48:44.35 & -44:32:31.42 &  & \\
 13.1 & 22:48:40.66 & -44:31:38.02 & [0.6-6.0] & 1.75$\pm$0.05\\
 13.2 & 22:48:41.82 & -44:32:13.59 &  & \\
 13.3 & 22:48:43.65 & -44:32:25.79 &  & \\
 14.1 & 22:48:43.21 & -44:32:18.35 & [0.6-6.0] & 1.02$\pm$0.02\\
 14.2 & 22:48:42.13 & -44:32:09.38 &  & \\
 14.3 & 22:48:41.26 & -44:31:48.90 &  & \\
\hline
\end{tabular}

\caption{As Table~\ref{tab:a2744-SL} but for Abell S1063.}
\end{table}

\begin{figure*}
\includegraphics[width=\textwidth]{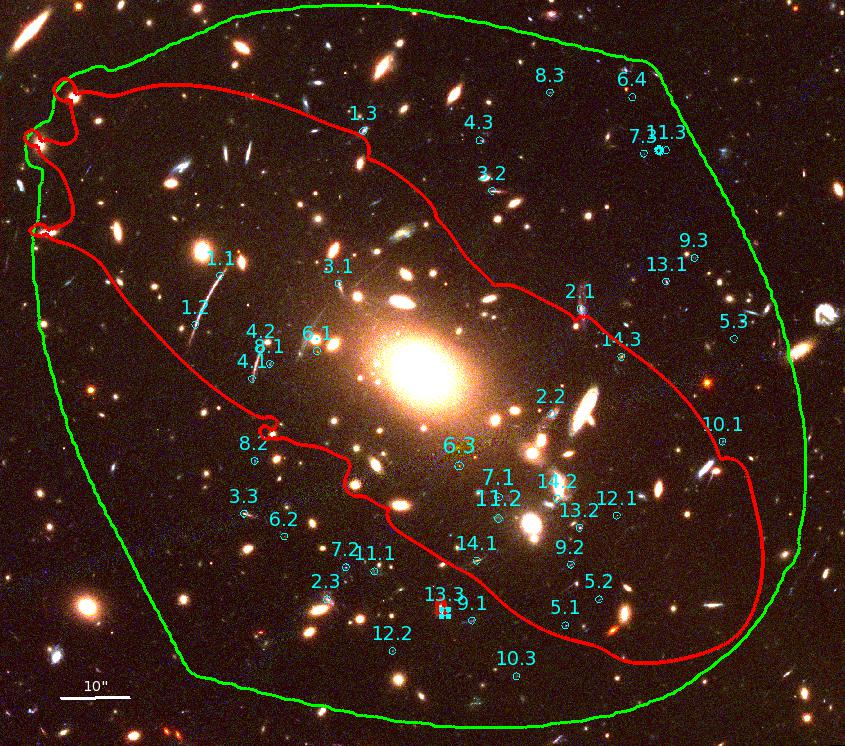}
\caption{\label{fig:r2248-imlarge}
Same as Fig.~\ref{fig:a2744-imlarge} but for Abell S1063 and filter combination  F606W+F814W+F160W.}
\end{figure*}

\subsubsection{Abell 370}

Our strong lensing analysis builds from the work published in \citet{Richard2010a}, where we identified 9 multiply imaged 
systems (as well as a faint radial arc, system 10). New HST/WFC3 images have been taken since this publication 
and allowed us to identify 2 additional systems. In total, we use 11 systems producing 34 images (Table \ref{tab:a370-SL}). 

\begin{table}
\begin{tabular}{lllll}
ID & $\alpha$ & $\delta$ & $z_{\rm prior}$ & $z_{\rm model}$ \\
   & [deg] & [deg] & & \\
\hline 
 1.1 & 02:39:52.10 & -01:34:37.28 & 0.806 & \\
 1.2 & 02:39:54.31 & -01:34:34.13 & 0.8060 & \\
 1.3 & 02:39:52.48 & -01:34:36.22 & 0.8060 & \\
 2.1 & 02:39:53.73 & -01:35:03.58 & 0.7250 & \\
 2.2 & 02:39:53.04 & -01:35:06.68 & 0.7250 & \\
 2.3 & 02:39:52.50 & -01:35:04.64 & 0.7250 & \\
 2.4 & 02:39:52.66 & -01:35:05.39 & 0.7250 & \\
 2.5 & 02:39:52.71 & -01:35:05.81 & 0.7250 & \\
 3.1 & 02:39:51.76 & -01:34:01.13 & [0.6-6.0] & 1.52$\pm$0.06\\
 3.2 & 02:39:52.45 & -01:33:57.38 &  & \\
 3.3 & 02:39:54.55 & -01:34:02.28 &  & \\
 4.1 & 02:39:55.12 & -01:34:35.41 & [0.6-6.0] & 1.34$\pm$0.03\\
 4.2 & 02:39:52.97 & -01:34:35.09 &  & \\
 4.3 & 02:39:50.87 & -01:34:40.97 &  & \\
 5.1 & 02:39:53.64 & -01:35:21.07 & [0.6-6.0] & 1.30$\pm$0.05\\
 5.2 & 02:39:53.08 & -01:35:21.67 &  & \\
 5.3 & 02:39:52.52 & -01:35:20.92 &  & \\
 6.1 & 02:39:52.68 & -01:34:38.32 & 1.063 & \\
 6.2 & 02:39:51.45 & -01:34:41.52 & 1.063 & \\
 6.3 & 02:39:55.12 & -01:34:38.12 & 1.063 & \\
 7.1 & 02:39:52.75 & -01:34:49.90 & [0.6-6.0] & 4.94$\pm$1.17\\
 7.2 & 02:39:52.77 & -01:34:51.09 &  & \\
 8.1 & 02:39:51.48 & -01:34:11.67 & [0.6-6.0] & 3.78$\pm$0.66\\
 8.2 & 02:39:50.86 & -01:34:25.67 &  & \\
 9.1 & 02:39:50.98 & -01:34:40.85 & [0.6-6.0] & 1.64$\pm$0.04\\
 9.2 & 02:39:52.68 & -01:34:34.94 &  & \\
 9.3 & 02:39:55.69 & -01:34:36.01 &  & \\
 11.1 & 02:39:51.32 & -01:34:10.12 & [0.6-6.0] & 5.93$\pm$0.15\\
 11.2 & 02:39:50.59 & -01:34:27.36 &  & \\
 12.1 & 02:39:52.74 & -01:34:00.43 & [0.6-6.0] & 4.59$\pm$0.44\\
 12.2 & 02:39:50.21 & -01:34:31.50 &  & \\
 12.3 & 02:39:56.19 & -01:34:15.70 &  & \\
 13.1 & 02:39:55.09 & -01:34:18.81 & [0.6-6.0] & 5.97$\pm$0.33\\
 13.2 & 02:39:54.05 & -01:34:08.02 &  & \\
\hline
\end{tabular}

\caption{\label{tab:a370-SL}
As Table~\ref{tab:a2744-SL} but for Abell 370.}
\end{table}

\begin{figure*}
\includegraphics[width=\textwidth]{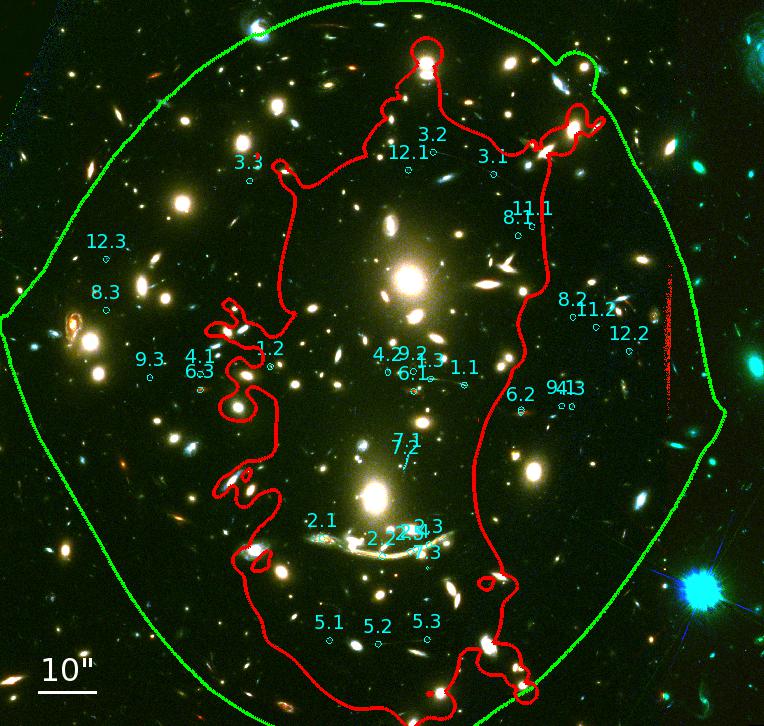}
\caption{\label{fig:a370-imlarge}
Same as Fig.~\ref{fig:a2744-imlarge} but for Abell 370 and filter combination  F625W+F814W+F160W.}
\end{figure*}

\subsection{Spectroscopy of strong-lensing features}

Whenever possible, we use spectroscopic redshifts of multiply imaged systems to constrain the lens model. This information is important 
since the source redshifts are degenerate with the absolute values of the model parameters characterising the mass distribution. All our models are anchored by spectroscopic redshifts for two to eight different multiple-image systems. We use spectroscopic redshifts from the literature, results shared among the different HFF mass modelling groups, and values obtained by us as part of dedicated spectroscopic follow-up observations. In the latter case we provide below the details of the observations, data reduction, and analysis of these spectra.

\subsubsection{A2744}

We targeted the core of this cluster as part of our VLT/FORS2 spectroscopic follow-up of submillimeter sources from the Herschel Lensing Survey \citep{Egami2010}. 10 multiple-image systems were observed in MOS mode for a total of 2 hrs exposure time during the night of September 3, 2011 (ESO program 087.B-0560(A), PI: Richard). We used a slit width of 1\arcsec, the GRIS\_300V grism and 
 the GG345 filter to cover the wavelength range from 4300 to 9200 \AA, albeit at low resolution (R$\sim$500-1000) and a dispersion of 2\AA\ per pixel. 
 
Of the strongly lensed sources targeted, redshifts could be secured for image A2744-4.3 at $z=3.58$, through strong Lyman-$\alpha$ emission (Fig. \ref{fig:specs}), in agreement with the photometric redshift of $z=3.5\pm0.3$ reported by \citet{Merten}. The spectrum of object 6.1 was found to feature a single narrow emission line at 5760\AA\ on top of a very blue continuum. The absence of other emission lines in the wavelength range covered agrees with $CIII]$ at $z=2.019$, which is compatible with our prediction based on a preliminary lens model.

\subsubsection{MACS\,J0416}
\label{spec-m0416}
We use spectroscopic redshifts for seven systems in MACS\,J0416.1$-$2403 (systems 2, 3, 7, 10, 13, 14, and 17) obtained as part of VLT programme 186.A-0798 (Balestra et al., Grillo et al., both in preparation), and shared among the different HFF mass modelling groups. We also include spectroscopic redshifts for system 1, published by our group in \citet{Christensen2012}.

\subsubsection{MACS\,J0717}

Since the analysis by \citet{2012A&A...544A..71L}, which presents our earlier observations on arc spectroscopy, 
one new spectroscopic redshift $z=6.4$ has been measured 
for system 19, as described previously. We use spectroscopic redshifts for six systems in total (systems 1, 3, 13, 14, 15 and 19), 
ranging from $z=1.850$ to $z=6.4$. 

\subsubsection{MACS\,J1149}

Since the analysis by \citet{Smith09}, no new spectroscopic redshifts has
been measured in MACS\,J1149.
We use spectroscopic redshifts for 3 systems (systems 1 to 3) at $z=1.490$, $z=1.894$ and $z=2.497$.
We refer to \citet{Smith09} for details regarding arc spectroscopy.

\subsubsection{AS1063}
\label{spec-as1063}

We have obtained spectroscopy of images 1.1 and 1.2 using multi-object spectroscopy with Magellan/LDSS3 on the night of October 11th 2009. 
We used LDSS3 with 1.0''-width slits and the VPH-All grism, which altogether provide a resolution of 650 and a dispersion of 1.9 \AA\ per pixel 
while covering the wavelength range 3800-9900 \AA\ . The seeing was good (0.7\arcsec-0.9\arcsec) during the 5.4 ksec total exposure time of these 
observations.
The LDSS3 spectra were calibrated, combined and extracted using standard IRAF and IDL routines. 
Both images AS1063-1.1 and 1.2 show a strong emission line at wavelength 8307 \AA\  and other fainter emission lines, 
compatible with $[OII]$ and $[NeIII]$ at  a spectroscopic redshift $z=1.229\pm0.005$. 

Additional spectroscopy was obtained on VLT/FORS2 the night of August 21st 2013, as part of the ESO program 291.A-5027 (PI:Richard). 
We designed a multi-slit spectroscopic mask covering images 2.1, 3.1, 4.3, 6.2, 6.3, 6.4 in the strong lensing region and we obtained 4.5 ksec 
of exposure time under bright time but good seeing conditions (0.8''-0.9''). We used the 600z grism, OG590 order filter and 1.0''-wide slits to cover the wavelength 
range 7200-9800 \AA\  at a resolution $\sim$1500-1900 and a dispersion 1.6 \AA\ per pixel. The FORS2 spectra were reduced using version 
4.9.11 of the FORS2 
data reduction software, and combined using standard IRAF and IDL routines. Both 2D and extracted spectra were visually inspected for faint 
emission lines and traces of continuum. 

\begin{figure*}\label{fig:specs}
\includegraphics[width=0.49\textwidth]{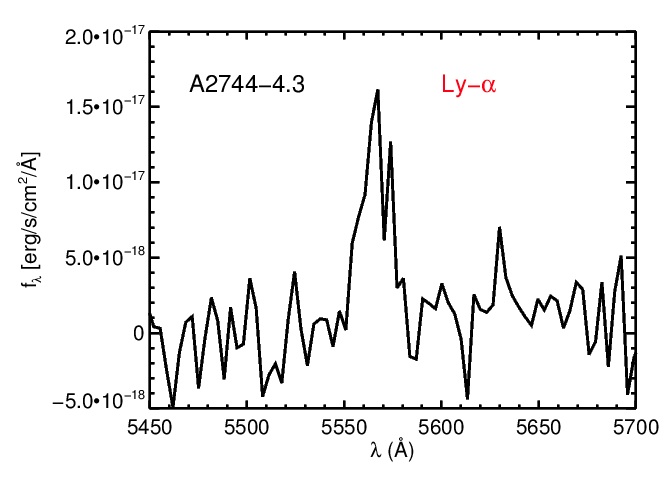}\includegraphics[width=0.49\textwidth]{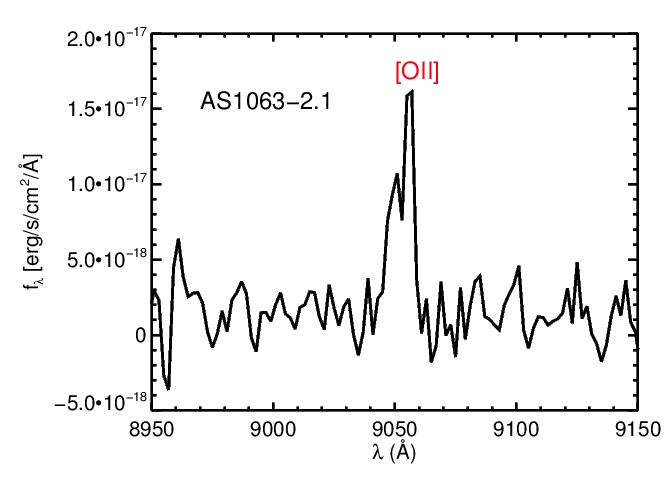}
\includegraphics[width=0.49\textwidth]{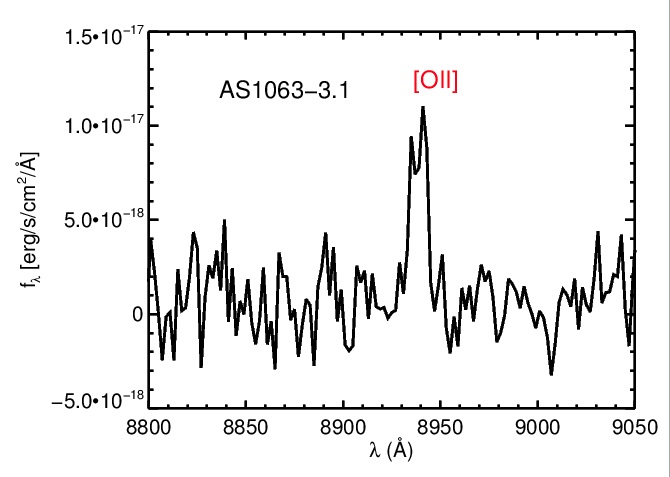}\includegraphics[width=0.49\textwidth]{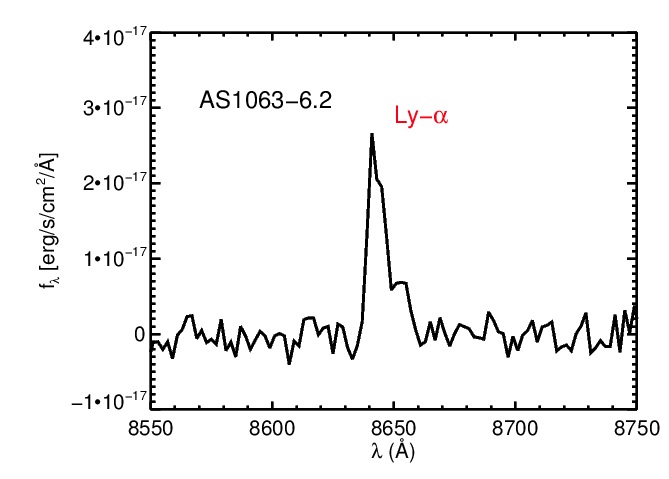}

\caption{Example of extracted spectra showing identified emission lines in multiple images.}
\end{figure*}

We identified a clear emission line doublet (also visible in single exposures) for images AS1063-2.1, 3.1 and 4.3, corresponding to $[OII]$ nebular line at 
$z=1.429$, $z=1.398$ and $z=1.260$, respectively (Fig. \ref{fig:specs}). All three slits covering system 6 (AS1063-6.2, 6.3 and 6.4) show an identical 
spectrum with a strong asymmetric emission line peaking at  $\lambda=8642$ \AA , which is the signature of Lyman-$\alpha$ emission at $z=6.107$ (Fig. \ref{fig:specs}). This confirms both the association of the 
three images as well as the high redshift nature of this source (see also \citet{Boone2013}).

All the redshifts measured in this cluster are in perfect agreement with estimations from a preliminary model of the mass distribution.

\subsubsection{Abell 370}

We targeted the core of this cluster in a VLT/FORS2 spectroscopic program meant to estimate cosmological parameters from strongly lensed features \citep{2010Sci...329..924J}. 9 multiply imaged systems were observed in MOS mode during a total of 7.5hrs exposure time in the nights between September 22 and 26 2011 (ESO program 087.A-0326(A), PI: Jullo). We have used 1"-width slits with the GRIS300V grism (2.5 hrs exposure time) to search for Ly-$\alpha$ emission or other UV features, and the GRIS600z grism and OG590 filter (5 hrs exposure time) to detect objects in the redshift range $1<z<2$. The dispersion per pixel is 1.63 \AA\ in the red, and 3.3 \AA\ in the blue.
Among the strongly lensed sources targeted, we were able to measure the redshifts of image A370-6.3 at $z=1.063$, through a clear detection of the $[OII]$ doublet. 
We also detect faint emission lines for image 3.1 and 3.2, which would correspond to $[OII]$ and $[NeIII]$ at $z=1.421$, and a faint emission line for image 4.1 at $z=1.275$ 
with the blue grism. The redshifts measured for systems 4 and 6 agree with the lens model predictions. However, the redshift of system 3 gives a very large $\chi^2$ to this system. Therefore, we preferred not to use the two most uncertain redshifts (systems 3 and 4) as 
constraints in our lens model (Table \ref{tab:a370-SL}). Additional follow-up spectroscopy would help us confirm these two redshifts.

\subsection{Weak Lensing constraints} 
\begin{table*}
\label{tab:wl-tab}
\begin{tabular}{lccccc}
Target & Color-Color selection & Density of background sources \\ \hline 
A2744 & F435W-F606W-F814W & 61 gal.arcmin$^{-2}$ \\ 
MACSJ0416.1-2403 & F475W-F625W-F814W & 50 gal.arcmin$^{-2}$  \\ 
MACSJ0717.5+3745 & F475W-F625W-F814W & 51 gal.arcmin$^{-2}$ \\ 
MACSJ1149.5+2223 & F475W-F625W-F814W & 61 gal.arcmin$^{-2}$  \\ 
AS1063 & F475W-F625W-F814W & 64 gal.arcmin$^{-2}$  \\ 
A370 & F475W-F625W-F814W & 74 gal.arcmin$^{-2}$  \\ \hline
\end{tabular}
\caption{Weak-lensing background galaxy densities obtained for each HFF cluster, as well as the HST-ACS filters used for the color-color selection to identify foreground galaxies and cluster members.}
\end{table*}

The background galaxy catalogues for the 6 clusters were derived following the method presented in \cite{Jauzac12} (hereafter J12). Therefore we here give a brief summary of the different steps.

The weak lensing analysis is based on shape measurements in the ACS/F814W band. Following a method developed for the analysis of data obtained for the COSMOS survey, and described in \cite{Leauthaud07} (hereafter L07), the SE\textsc{xtractor} photometry package \citep{1996A&AS..117..393B} is used for the detection of the sources, using the \emph{'Hot-Cold'} method (\citealt{rix04},L07). This detection configuration combines an optimal detection of the brightest objects (\emph{cold} step), and of the faintest ones (\emph{hot} step). The resulting catalogue is then cleaned by removing spurious, duplicate detections, and any sources in the vicinity of stars or saturated pixels. The star-galaxy classification is performed using a standard MAG$\_$AUTO-MU$\_$MAX plane selection (see L07 \& J12 for more details). Finally, to overcome the pattern-dependent correlations introduced by the drizzling process between neighboring pixels, which artificially reduce the noise level of co-added drizzled images, we apply the remedy used by L07: simply scaling up the noise level in each pixel by the same constant $F_{A} \approx 0.316$, defined by \cite{Casertano00}.

Since only galaxies behind the clusters are gravitationally lensed, the presence of cluster members dilutes the observed shear and reduces the significance of all quantities derived from it (see J12 for a more detailed discussion). Therefore, the identification and the removal of the contaminating unlensed galaxies is crucial.  
Thanks to the existing HST data for all 6 clusters, a minimum of two colours is available for each of them. Therefore, following the methodology described in J12, we used the existing spectroscopic and photometric redshifts \citep{2011ApJ...728...27O,2014ApJS..211...21E} to calibrate a color-color selection to identify the foreground galaxies and cluster members (the combination of filters used for each cluster is given in Tab.~\ref{tab:wl-tab}). 

The measure of galaxy shapes is done using the RRG method \citep{Rhodes00}, which was developed for the analysis of data obtained from space, and is thus ideally suited for use with a small, diffraction-limited PSF as it decreases the noise on the shear estimators by correcting each moment of the PSF linearly, and only dividing them at the very end to compute an ellipticity.
The last step of the weak lensing catalogue construction consists of applying lensing cuts, i.e., to exclude galaxies whose shape parameters are ill-determined, and will thus increase the noise in the shear measurements more than add to the shear signal. More details can be found in J12. Finally, we assume the redshift distribution of \citet{1995ApJ...455L..99S} for the background galaxies with shear measurement. The final density of background galaxies we obtained for each clusters are given in Tab.~\ref{tab:wl-tab}.

\section{Methodology}
\label{sec-sl}

We combine strong-lensing and weak-lensing constraints to model the mass distribution in each cluster. Following our successful strategy from previous lensing work (e.g., \citealt{2007ApJ...668..643L,2012A&A...544A..71L,Richard2010a,2010MNRAS.404..325R}), we adopt a parametric model combining both large-scale (cluster or group size) mass clumps as well as galaxy-scale mass clumps. In the following, we describe the selection of cluster members across the ACS field of view, the choice of model parameters, and their optimisation.

\subsection{Galaxy catalogues}

We create object catalogues for each HFF cluster using SExtractor \citep{1996A&AS..117..393B} for the areas shown in Fig.~\ref{fig:hff-layout}, i.e., the region within which \textit{HST} imaging data are available in at least three well separated passbands. Using the F606W images (F625W in the case of A370) as the primary detection band, we run SExtractor in dual-image mode on the images in each passband. The resulting catalogues are combined to create for each cluster a master catalogue of objects. Star-galaxy separation is performed by identifying stars in two separate parameter spaces, namely peak surface brightness vs.\ flux, and half-light radius vs.\ flux. We remove from our catalogues all stars and all spurious sources found either to feature a higher peak surface brightness or to be more compact than stars.  To improve the robustness of this procedure (the small area covered by the existing \textit{HST} images of the HFF contains very few stars for each individual field), we perform the star-galaxy separation on a combined object catalogue for all six fields. 

\begin{figure}
\includegraphics[width=0.5\textwidth]{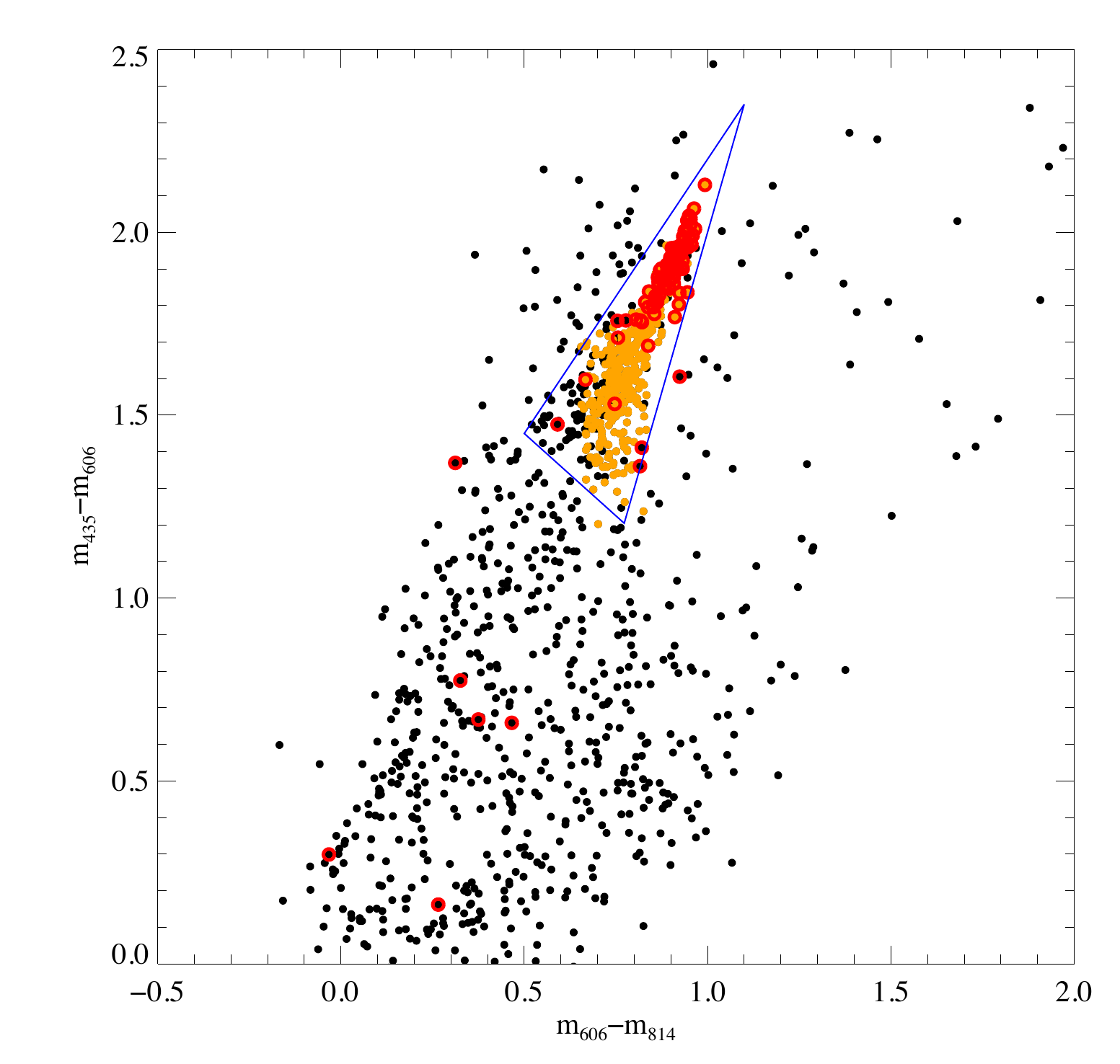}
\caption{Colour-colour diagram for galaxies in the A2744 field, as defined in Fig.~\ref{fig:hff-layout}. Galaxies marked in orange meet the dual red-sequence criterion, i.e., they fall within $3\sigma$ of the cluster red sequence in both the  ($m_{\rm F606W}{-}m_{\rm F814W}$) vs $m_{\rm F814W}$ and the ($m_{\rm F435W}{-}m_{\rm F606W}$) vs $m_{\rm F814W}$ colour-magnitude diagrams. Galaxies highlighted in red are spectroscopically confirmed cluster members. \label{fig:a2744cc}}
\end{figure}

The resulting catalogues of galaxies are then used to establish colour-colour cuts for the identification of likely cluster members. We examine the distribution of galaxies in both colour-magnitude diagrams and in colour-colour space ($m_{\rm F435W}{-}m_{\rm F606W}$ vs $m_{\rm F606W}{-}m_{\rm F814W}$), highlighting the loci of spectroscopically confirmed cluster members. We then define probable cluster members to be those galaxies that fall within $3\sigma$ of a linear model of the cluster red sequence in both the ($m_{\rm F606W}{-}m_{\rm F814W}$) vs $m_{\rm F814W}$ and the ($m_{\rm F435W}{-}m_{\rm F606W}$) vs $m_{\rm F814W}$ colour-magnitude diagrams. Fig.~\ref{fig:a2744cc} shows the galaxies selected by this process as well as all spectroscopically confirmed cluster members. We compile the latter from various catalogues of spectroscopic redshifts: for A2744, we consult \citet{2011ApJ...728...27O}; for MACS\,J0416, MACS\,J0717, and MACS\,J1149, we use the redshifts published by Ebeling, Ma \& Barrett (2013); and for AS1063 and A370 we rely on spectroscopic redshifts compiled in the NASA Extragalactic Database (NED). We also use redshifts obtained by Balestra et al. (2013, in prep) for MACS\,J0416. As shown in Fig.~\ref{fig:a2744cc}, the galaxies selected by the dual red-sequence criterion fall into a well defined triangular region of ($m_{\rm F435W}{-}m_{\rm F606W}$) vs ($m_{\rm F606W}{-}m_{\rm F814W}$) colour-colour space. Galaxies featuring luminosities exceeding $M_{\rm F814W}=0.01\, L^\ast$ are included in our strong-lens mass model as small-scale perturbers (see following section).

\subsection{Model parametrization}
  
Producing a magnification map involves solving the lens equation for light rays originating from distant sources and deflected by the massive foreground cluster. This is ultimately an inversion problem for which several sets of codes and approaches have been developed independently. Our collaboration uses Lenstool\footnote{publically available \href{http://projects.lam.fr/projects/lenstool/wiki}{on the dedicated web page}}  \citep{2007NJPh....9..447J}, an algorithm we developed collectively over the years. In this software, the cluster mass distribution can be described as a combination of physically motivated mass components, both for the 
individual galaxies and the smoother, large-scale haloes. 

As a basis function for our models, we adopt the dual pseudo-isothermal elliptical mass distribution (dPIE, also known as truncated PIEMD),  which corresponds to an isothermal profile with two characteristic radii: a core radius r$_{\rm core}$ (producing a flattening of the mass distribution at the centre), and a cut radius r$_{\rm cut}$ (producing a drop-off of the mass distribution on large scales). More details on the dPIE parametrization are given in \citet{Richard09} and \citet{2012A&A...544A..71L}.

This method has the advantage that the geometrical parameters of the galaxy-scale components (centre, ellipticity, orientation)  can be directly related to the shape parameters measured from the light distribution of cluster galaxies in our photometric catalog. In order to limit the number of free parameters in our model, we use the F814W band as the reference band for these shape measurements and determine the three other parameters of each galaxy's dPIE description (central velocity dispersion $\sigma$, core and cut radii) from scaling relations based on the galaxy's luminosity $L$ relative to $L^*$. We set $r_{\rm core}^*$=0.15 kpc for all models, but optimise both $\sigma^{*}$ and $r_{\rm cut}^*$, following \citet{Smith09} and \citet{2012A&A...544A..71L}, respectively.  We adopt a flat prior for $r_{\rm cut}^*$ in the range $[1-100]$ kpc, and a Gaussian prior $\sigma^{*}=158\pm27$ km/s, as demonstrated by \citet{2010MNRAS.404..325R} for a sample of $z\sim0.2$ clusters.

 At present, the prevailing modeling approach is to assign a small-scale dark-matter clump to each cluster galaxy in our catalog, and a large-scale dark-matter clump to prominent concentrations of cluster galaxies. This technique has proven very reliable and provides results in mass distributions in reasonably good agreement with theoretical predictions from high-resolution cosmological N-body simulations \citep{NatarajanKneib1997,Natarajan2007}. This explicit one-to-one correspondence between mass and light is less accurate, however, in the outer regions of clusters where the galaxy distribution is sparser, and strong-lensing constraints are unavailable. The solution is to combine strong-lensing constraints near the cluster cores with a weak-lensing analysis on larger scales.
  
\subsection{Likelihoods definition}
\label{sec-slwl}
  
For each Frontier Field cluster, we optimise two parametric models: an SL (strong lensing) model solely based on the strong-lensing constraints presented in Sect. \ref{sec-multiples}, and an SL+WL model combining the strong-lensing constraints with a weak-lensing analysis, as described in Sect. \ref{sec-slwl}. We proceed in two steps for each cluster: we start by creating an SL model, optimised in the image plane, and determine the parametrisation (number of clumps and individual galaxies to be optimised). In this case, we used the following likelihood definition

\begin{equation}
\mathcal{L}_{SLimg} = \prod^{M,N_i}_{i,j} \frac{1}{\sqrt{2 \pi \sigma_{SL}^2}} \exp \left( -\frac{1}{2} \frac{|\theta_{ij} - \langle \theta_i \rangle|^2}{\sigma_{SL}^2}  \right)
\end{equation}

\noindent where $\langle \theta_i \rangle$ is the estimated image position for system $i$, based on the barycenter of the multiple positions in the source plane. 
 
We then use the same parametrisation to optimise the SL+WL model, this time performing the optimisation in the source plane which is less computing-time intensive. To this end, we define the total likelihood $\mathcal{L} = \mathcal{L}_{SLsrc}  \times \mathcal{L}_{WL}$ as the product of the strong- and the weak-lensing likelihoods. The strong-lensing likelihood in the source plane is, in turn, defined as the product of $M$ systems of $N_i$ multiple-image likelihoods

\begin{equation}
\mathcal{L}_{SLsrc} = \prod^{M,N_i}_{i,j} \frac{1}{\sqrt{2 \pi \mu_{ij}^{-1} \sigma_{SL}^2}} \exp \left( -\frac{1}{2} \frac{|\beta_{ij} - \langle \beta_i \rangle|^2}{\mu_{ij}^{-1} \sigma_{SL}^2}  \right)
\end{equation}

\noindent where $\beta_{ij}$ are the source positions of the multiple images,  and $\langle \beta_i \rangle$ are the barycenter of these positions for system $i$. For each image, the positional uncertainty in the image-plane $\sigma_{SL}$ is multiplied by the amplification $\mu_{ij}$. We conduct this optimisation in the source plane as the much more involved computations of an image-plane optimisation would have unduly strained the available computing resources. We find that, for strong-lensing datasets only, source- and image-plane optimisations yield similar reconstructions for the clusters studied in this work.

Finally, the weak-lensing likelihood is defined as the product of the $L$ weak-lensing source likelihoods

\begin{equation}
\mathcal{L}_{WL} = \prod^{L}_i \frac{1}{\sqrt{2 \pi \sigma_{\epsilon_i}^2}}Ê\exp \left( -\frac{1}{2} \frac{|\epsilon^s_i|^2}{\sigma_{\epsilon_i}^2} \right)
\end{equation}

\noindent where $|\epsilon^s_i|$ is the module of the predicted source ellipticity obtained from the amplification matrix $\mathcal{A}$ and the second brightness moments of each image $\mathcal{Q}$, through the equation $\mathcal{Q}^s_i = \mathcal{A} \mathcal{Q} \mathcal{A}^T$ \citep{2001ASPC..237..421B}. The ellipticity is defined as $e = (a^2 - b^2) / (a^2 + b^2)$, where $a$ and $b$ are the eigenvalues of $\mathcal{Q}$. This matrix transformation is valid both in the weak and in the strong lensing regime. In this work, we assume $\sigma_{\epsilon_i}^2 = \sigma_{\rm int}^2 + \sigma_{\rm meas_i}^2$, i.e., the variance is the quadratic sum of the intrinsic ellipticity and the shape measurement errors for each galaxy.

By adopting this two-step approach, we avoid the possibility of reaching a local minimum in the source-plane optimisation: in the SL+WL model group- and cluster-scale haloes are mainly optimised on large scales based on weak-lensing constraints, while their center and shape parameters remain similar to those of the SL model.

\section{Results}

\subsection{Parametric model}

The number of large-scale clumps included in each cluster was chosen to minimize the number of free parameters while reproducing the multiple-image sets to an accuracy of better than 0.8$\arcsec$. The number of large-scale clumps finally adopted for our models varies from 1 (for Abell S1063) to 5 (for Abell 2744 and MACS\,J0717). As shown in Sect. \ref{sec-multiples}, Abell 2744 features a bimodal mass distribution on large scales, but also includes a number of group-scale structures at the edges of the ACS field of view.  For the even more complex system MACS\,J0717 the distribution of cluster light shows peaks at the location of each large-scale clump (see the discussion in \citealt{2012A&A...544A..71L}). 

In addition to these large-scale haloes, a few individual galaxies are explicitly included in our lens models, as their adopted properties have a strong influence on the location of nearby multiple-imaged systems. This is true for two galaxies in Abell 2744,  two galaxies in Abell 370 (as already found by \citealt{Richard2010a}), as well as for a massive foreground galaxy at $z\sim0.1$ in the field of MACS\,J0416 which does not follow the same scaling relations as the cluster members.

The resulting parameters of the combined (SL+WL) models are summarised in Table \ref{tab:resultsSLWL}.  For each cluster we describe the large-scale potentials (identified as DM1 to DM5), the aforementioned added galaxy-scale potentials (GAL1 and GAL2), and the scaling relations for cluster members, presented for a L$^*$ galaxy. The optimised source redshift for all multiple-image systems without spectroscopic redshift is summarised in the last column of Tables \ref{tab:a2744-SL} to \ref{tab:a370-SL}. The parameters of the combined models agree with those of the best-fit  SL models within the 2$\sigma$ uncertainties; the  resulting $\chi2$ values are very similar too.  As expected, the most massive components detected by our weak-lensing analysis were thus already present in our SL parametrisation.

%
%

\begin{table*}
\begin{tabular}{llllllll}
Potential & $\Delta\alpha$ & $\Delta\delta$ & $e$ & $\theta$ & r$_{\rm core}$ & r$_{\rm cut}$ & $\sigma$ \\
   & [arcsec] & [arcsec] & & [deg] & [kpc] & [kpc] & [km/s] \\
\hline
\multicolumn{6}{l}{Abell 2744, $\alpha$ = 00:14:20.698, $\delta$ = -30:24:00.60, $z=0.308$}\\
\hline
DM1 & $ -7.0^{+  1.0}_{ -0.9}$ & $ -6.4^{+  1.3}_{ -1.3}$ & $ 0.21^{+ 0.04}_{-0.05}$ & $ 82.9^{+  3.8}_{ -7.9}$ & $107^{+12}_{-9}$ & $[1000]$ & $826^{+42}_{-56}$ \\
DM2 & $-17.9^{+  0.3}_{ -0.4}$ & $-18.1^{+  0.4}_{ -0.3}$ & $ 0.57^{+ 0.04}_{-0.04}$ & $ 43.0^{+  2.1}_{ -1.3}$ & $35^{+3}_{-2}$ & $[1000]$ & $621^{+26}_{-17}$ \\
DM3 & $[ 24.2]$ & $[155.8]$ & $[0.30]$ & $[-74.8]$ & $116^{+0}_{-15}$ & $[1000]$ & $665^{+30}_{-50}$ \\
DM4 & $104.4^{+  2.4}_{ -1.1}$ & $ 83.1^{+  0.4}_{ -1.8}$ & $ 0.16^{+ 0.07}_{-0.08}$ & $  8.9^{+  6.9}_{ -3.9}$ & $64^{+8}_{-6}$ & $[1000]$ & $747^{+42}_{-34}$ \\
DM5 & $  0.2^{+  1.0}_{ -1.1}$ & $ 20.3^{+  1.0}_{ -1.3}$ & $ 0.67^{+ 0.01}_{-0.13}$ & $126.0^{+  9.9}_{ -9.0}$ & $37^{+4}_{-3}$ & $[1000]$ & $439^{+46}_{-22}$ \\
GAL1 & $[-12.7]$ & $[ -0.8]$ & $[0.30]$ & $[-46.6]$ & $[0]$ & $[43]$ & $6^{+29}_{-56}$ \\
GAL2 & $[ -3.6]$ & $[ 24.7]$ & $[0.72]$ & $[-33.0]$ & $[0]$ & $[34]$ & $153^{+24}_{-69}$ \\
L$^{*}$ galaxy &  & & & & $[0.15]$ & $7^{+2}_{-2}$ & $167^{+5}_{-8}$\\
\hline

\multicolumn{6}{l}{MACS\,J0416 $\alpha$ = 04:16:09.144, $\delta$ = -24:04:02.95, $z=0.396$}\\
\hline
DM1 & $ -5.6^{+  1.0}_{ -0.6}$ & $  2.7^{+  0.7}_{ -0.7}$ & $ 0.71^{+ 0.04}_{-0.06}$ & $146.5^{+  1.5}_{ -1.7}$ & $76^{+12}_{-7}$ & $[1000]$ & $809^{+46}_{-38}$ \\
DM2 & $ 23.7^{+  1.4}_{ -0.7}$ & $-45.7^{+  1.5}_{ -1.4}$ & $ 0.56^{+ 0.04}_{-0.06}$ & $128.5^{+  0.8}_{ -0.6}$ & $120^{+10}_{-7}$ & $[1000]$ & $1019^{+36}_{-43}$ \\
GAL1 & $[ 31.8]$ & $[-65.5]$ & $[0.04]$ & $[-40.4]$ & $[0]$ & $[62]$ & $[140]$ \\
L$^{*}$ galaxy &  & & & & $[0.15]$ & $9^{+15}_{-8}$ & $183^{+16}_{-16}$\\
\hline

\multicolumn{6}{l}{MACS\,J0717 $\alpha$ = 07:17:35.575, $\delta$ = +37:44:44.57, $z=0.545$}\\
\hline
DM1 & $  5.0^{+  1.8}_{ -1.2}$ & $ 16.2^{+  2.3}_{ -3.1}$ & $ 0.59^{+ 0.09}_{-0.08}$ & $ 68.9^{+  6.1}_{ -8.3}$ & $34^{+29}_{-5}$ & $[1000]$ & $837^{+56}_{-58}$ \\
DM2 & $ 35.5^{+  2.6}_{ -1.8}$ & $-10.1^{+  3.1}_{ -3.5}$ & $ 0.95^{+ 0.04}_{-0.09}$ & $ 49.4^{+  4.8}_{ -4.2}$ & $60^{+31}_{-18}$ & $[1000]$ & $719^{+89}_{-67}$ \\
DM3 & $ 71.1^{+  4.7}_{-11.3}$ & $ 35.8^{+  3.1}_{ -4.1}$ & $ 0.90^{+ 0.03}_{-0.04}$ & $ 20.1^{+  4.5}_{ -3.0}$ & $154^{+49}_{-26}$ & $[1000]$ & $1082^{+70}_{-209}$ \\
DM4 & $ 98.1^{+  5.7}_{ -9.3}$ & $ 72.0^{+  5.9}_{ -5.5}$ & $ 0.81^{+ 0.13}_{-0.12}$ & $-24.1^{+ 24.4}_{ -7.0}$ & $35^{+49}_{-28}$ & $[1000]$ & $521^{+138}_{-83}$ \\
DM5 & $[-19.4]$ & $[-21.7]$ & $[0.23]$ & $[-40.0]$ & $[2]$ & $[392]$ & $[180]$ \\
L$^{*}$ galaxy &  & & & & $[0.15]$ & $67^{+12}_{-18}$ & $135^{+25}_{-24}$\\
\hline

\multicolumn{6}{l}{MACS\,J1149 $\alpha$ = 11:49:35.695, $\delta$ = +22:23:54.70, $z=0.544$}\\
\hline
DM1 & $ -0.8^{+  1.4}_{ -2.3}$ & $  1.4^{+  1.5}_{ -1.6}$ & $ 0.63^{+ 0.05}_{-0.07}$ & $ 35.1^{+  2.0}_{ -2.5}$ & $201^{+61}_{-10}$ & $[1000]$ & $1242^{+84}_{-84}$ \\
DM2 & $-23.1^{+  1.4}_{ -2.5}$ & $-23.7^{+  1.1}_{ -2.4}$ & $[0.00]$ & $[ 34.0]$ & $3^{+13}_{-1}$ & $[1000]$ & $235^{+56}_{-32}$ \\
DM3 & $  9.6^{+  3.9}_{ -1.0}$ & $ 40.3^{+  1.3}_{ -0.1}$ & $[0.00]$ & $[ 34.0]$ & $11^{+28}_{-3}$ & $[1000]$ & $407^{+59}_{-48}$ \\
DM4 & $-13.6^{+  2.2}_{ -1.4}$ & $ 98.3^{+  1.7}_{ -1.9}$ & $[0.23]$ & $[-66.2]$ & $21^{+14}_{-6}$ & $[1000]$ & $363^{+76}_{-34}$ \\
DM5 & $[  0.0]$ & $[  0.0]$ & $[0.20]$ & $[ 34.0]$ & $17^{+12}_{-5}$ & $230^{+109}_{-38}$ & $445^{+5}_{-62}$ \\
L$^{*}$ galaxy &  & & & & $[0.15]$ & $30^{+11}_{0}$ & $153^{+3}_{-1}$\\
\hline

\multicolumn{6}{l}{AbellS 1063 $\alpha$ = 22:48:43.973, $\delta$ = -44:31:51.20, $z=0.348$}\\ 
\hline
DM1 & $  0.8^{+  0.4}_{ -0.3}$ & $  0.1^{+  0.3}_{ -0.3}$ & $ 0.58^{+ 0.01}_{-0.01}$ & $-37.3^{+  0.2}_{ -0.2}$ & $120^{+3}_{-4}$ & $[1000]$ & $1374^{+9}_{-7}$ \\
L$^{*}$ galaxy &  & & & & $[0.15]$ & $32^{+12}_{-2}$ & $104^{+18}_{-24}$\\
\hline

\multicolumn{6}{l}{Abell 370 $\alpha$ = 02:39:53.076, $\delta$ = -01:34:56.14, $z=0.375$}\\
\hline
DM1 & $  3.1^{+  0.5}_{ -0.5}$ & $  8.6^{+  0.1}_{ -0.6}$ & $ 0.59^{+ 0.04}_{-0.04}$ & $-106.0^{+  2.8}_{ -3.3}$ & $64^{+8}_{-5}$ & $[1000]$ & $833^{+58}_{-6}$ \\
DM2 & $ -2.5^{+  0.8}_{ -0.0}$ & $ 35.6^{+  1.5}_{ -2.2}$ & $ 0.38^{+ 0.04}_{-0.05}$ & $-89.6^{+  2.8}_{ -2.4}$ & $155^{+9}_{-12}$ & $[1000]$ & $1128^{+37}_{-51}$ \\
GAL1 & $[ -0.0]$ & $[  0.0]$ & $[0.30]$ & $[-81.9]$ & $[0]$ & $34^{+4}_{-3}$ & $129^{+22}_{-22}$ \\
GAL2 & $[  7.9]$ & $[ -9.8]$ & $[0.26]$ & $[ 25.7]$ & $[0]$ & $28^{+5}_{-4}$ & $64^{+11}_{-16}$ \\
L$^{*}$ galaxy &  & & & & $[0.15]$ & $61^{+21}_{-5}$ & $116^{+16}_{-8}$\\
\hline

\end{tabular}
\caption{
\label{tab:resultsSLWL}
Best fit parameters of the mass components in each cluster for the models optimised with the combination of strong-lensing and 
weak-lensing constraints. From left to right: identification of potential (DM: cluster-scale dark-matter halo, GAL: galaxy-scale halo, 
L$^{*}$: scaling relation parameters for L$^{*}$ galaxy in cluster members), relative astrometric position of center of potential, ellipticity 
and position angle, core and cut radii, velocity dispersion.}
\end{table*}

\subsection{Output maps and error estimation}

The parametric models are adjusted with Lenstool in a Bayesian way, i.e., we probe their posterior probability density with a MCMC sampler \citep{2007NJPh....9..447J}. This process allows us to easily and reliably estimate the errors on derived quantities such as the amplification maps and the mass maps. For each cluster, we use 200 randomly selected models to sample the posterior-probability distribution of each parameter.

\begin{figure}
\includegraphics[width=0.5\textwidth]{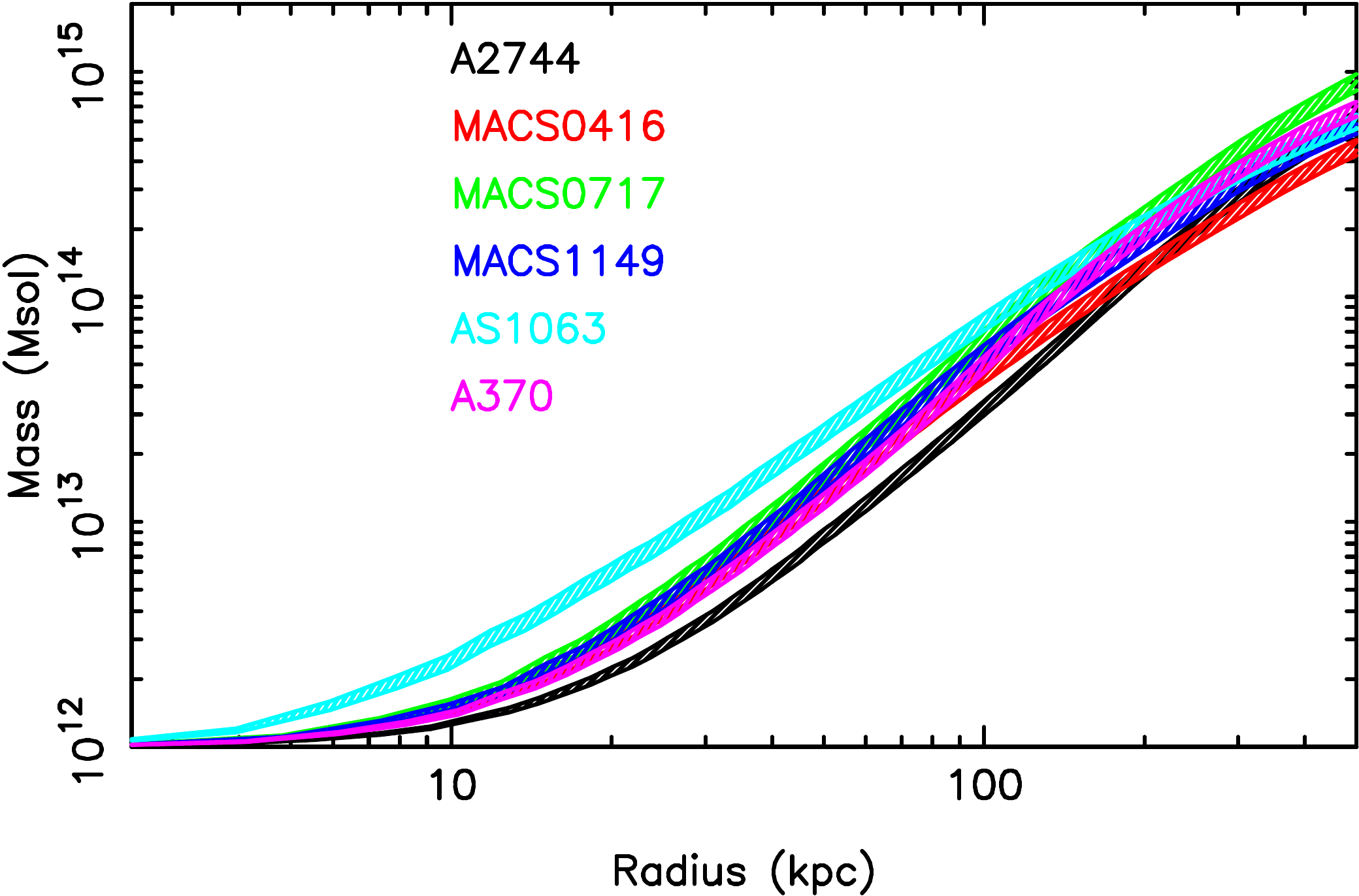}
\includegraphics[width=0.5\textwidth]{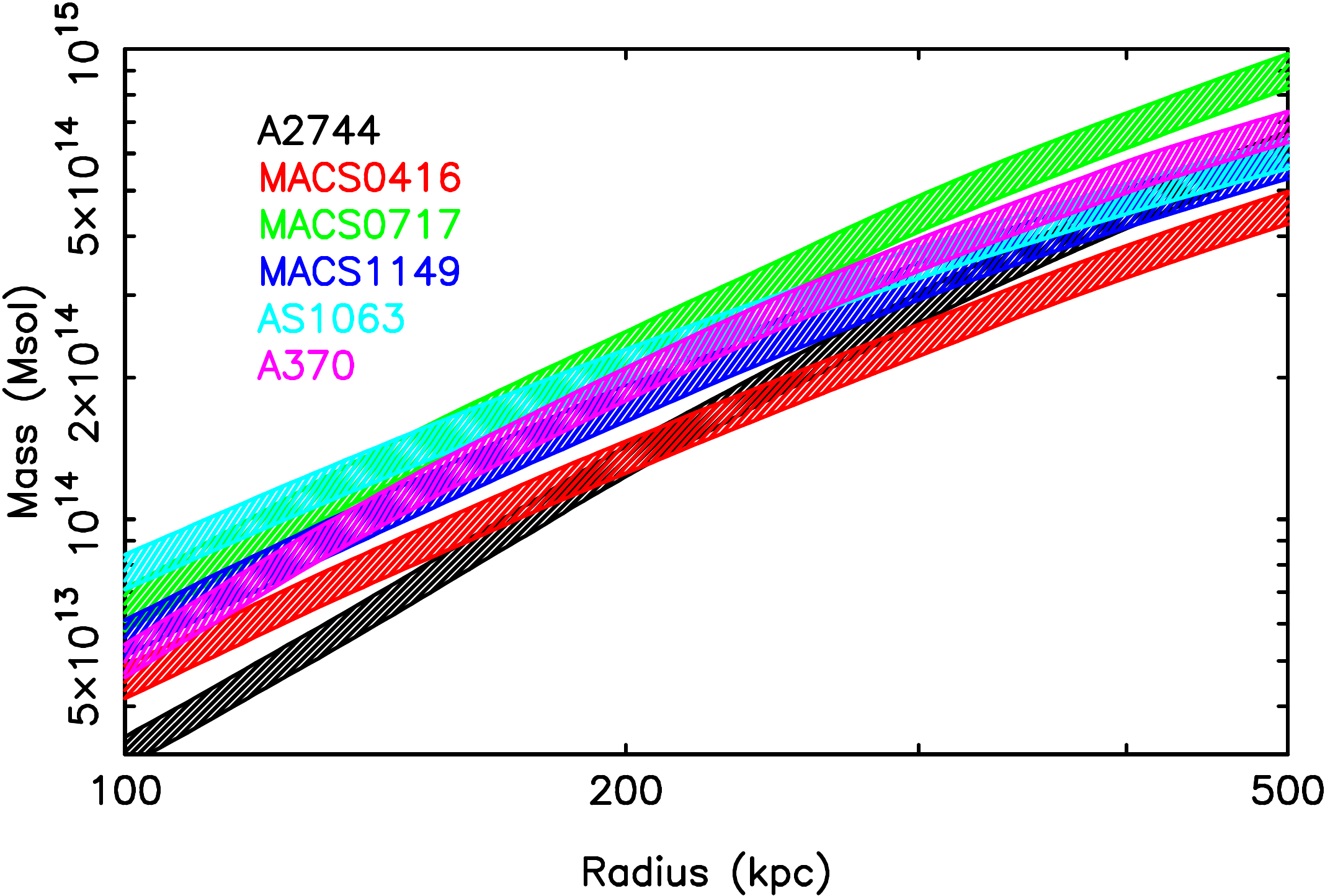}
\caption{\label{mprofiles} Total integrated mass  as a function of the projected 
distance from the barycenter. Each color hatched region corresponds to the average and 1 $\sigma$ 
dispersion on the integrated mass in each FF cluster. The bottom panel shows a zoom over the $100<r<500$ kpc 
region.}
\end{figure}

High-resolution mass maps (integrated over the line of sight) were produced for each of these models, and then integrated as a function of the radial distance from their barycenter. The resulting average integrated mass and 1$\sigma$ dispersion (computed over the 200 maps per cluster) are presented in Figure \ref{mprofiles}. All FF clusters are found to be very massive, reaching  integrated masses at 500 kpc radius between 4$\times$10$^{14}$ (for MACS\,J0416) and 10$^{15}$ M$_\odot$ for the most massive cluster MACS\,J0717. At $r<100$ kpc, all six mass profiles look very similar, the sole exception being the highly concentrated and fully relaxed system Abell S1063, whose brightest cluster galaxy is almost exactly centred within a single large-scale dark-matter halo. 
 
\begin{figure*}
\includegraphics[width=\textwidth]{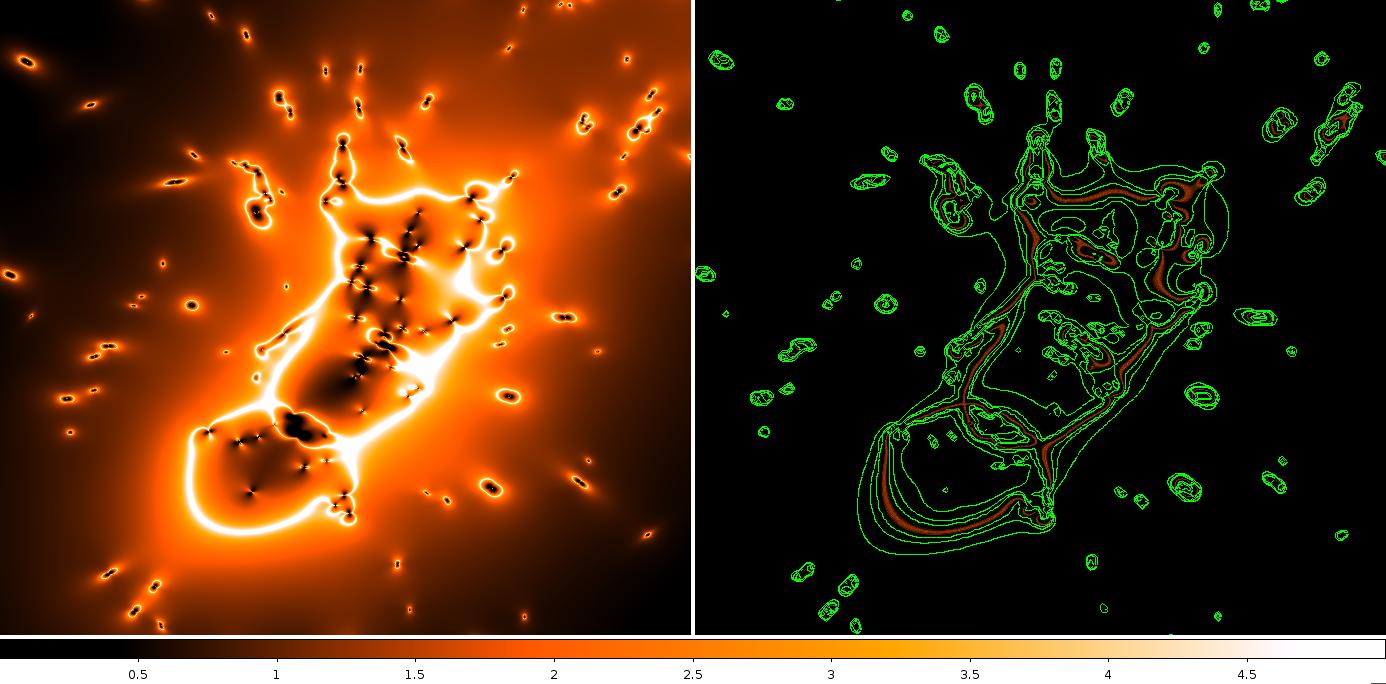}
\caption{\label{fig:muerrors}Magnification map (left panel) and error map (right panel) produced in the central region of Abell 2744 for a source at $z=9$. Contours show a relative error of 10,20 and 50\%}
\end{figure*}

We applied the same procedure to  create amplification maps and accompanying error maps in all six fields, assuming fiducial source redshifts of $z=1$, $z=2$, $z=4$ or $z=9$.  Examples are shown in Figure \ref{fig:muerrors}. These maps can also be extrapolated to the locations of the blank fields for the benefit of the larger extragalactic community.  Amplification maps at other source redshifts can be derived based on the convergence $\kappa$ and shear $\gamma$ maps before normalisation by the geometrical distance ratio between the cluster and the source. For a given image position and source redshift, one can derive the magnification $\mu$ as follows:

\begin{equation}
1/\mu=(1-(D_{\rm LS}/D_{\rm S})\ \kappa)^2-((D_{\rm LS}/D_{\rm S})\gamma)^2
\end{equation}

where DLS and DS are the angular diameter distance between the lens and the source, and between the observer and the source, 
respectively.

The best-fit lenstool models, mass maps, amplification maps and relative errors, as well as the 200 convergence and shear maps 
for each cluster, are made publicly available on the frontier fields website \url{http://archive.stsci.edu/prepds/frontier/lensmodels/}.

It is worth mentioning that the relative errors computed from the MCMC samples are only statistical errors and do not include 
systematics due to the assumptions made in the models. In order to estimate the level of these systematics we have performed 
three specific tests, where we re-optimised the six mass models under different assumptions : (a) we selected only 10 robust 
multiple systems per cluster, i.e. either spectroscopically confirmed and/or among the brightest systems in agreement with various lensing 
groups, as strong lensing constraints ; (b) we increased the error measurement on the shear values by 50\% ; and (c) we increased 
by $\Delta z$=0.5 the distribution assumed for background sources in the weak lensing constraints. By comparing the magnification 
maps at very high redshift, we observed an average variation in magnification within the ACS field-of-view by 0.1 to 0.35 magnitudes 
for test (a), by 0.1 magnitudes for test (b) and by 0.05 to 0.1 magnitudes for test (c), depending on the complexity of the cluster model. 
As expected due to its strong influence on high magnification values, the robustness of multiple systems used as constraints produces the 
stronger systematics. This justifies the need for future comparison between the magnification maps produced by the different modelling 
teams on the same simulated cluster.

\section{Discussion}

One of the key goals of the Frontier Fields initiative is to improve the statistics on faint distant galaxies observed during the epoch of reionization.  The expected extent of this improvement can be assessed using our best models for each cluster, which yield the gravitational magnification for high-redshift galaxies in the central region covered by deep observations with both ACS and WFC3.  In doing so, we adopt $z=7$ as a reference source redshift, noting though that the change in magnification is typically very small between $z=7$ and $z=10$. 

A first estimate of the relative lensing efficiency of the six HFF clusters can be obtained by comparing their peak magnification values. These are reached close to the \textit{critical lines}, which are shown in red in Fig. \ref{fig:a2744-imlarge} to Fig. \ref{fig:a370-imlarge}. The solid angles over which magnifications of $\mu>5$ and $\mu>10$ are attained in each cluster range from 0.8 to 2.9 arcmin$^2$ and are summarized in Table \ref{tab:numbers_highz}.  While the largest areas with significant magnification are provided by MACS\,J0717, Abell 370 and Abell S1063, all FF clusters are similarly efficient lenses to within a factor of two.

One shortcoming of peak magnification as a metric for measuring lensing efficiency is that it provides no information about image multiplicity, i.e., the frequency of the various lensing configurations (the creation of 1, 3, 5, or more images from a single source), which can vary from cluster to cluster. We therefore compute a second important cluster attribute, namely the sky area covered by multiple images in the image plane. This area is related to the number of multiple images expected to be found in each cluster.  The sky area within which multiple images are observed encloses the critical line but has a more circular shape, covering between 1.5 and 4.2 arcmin$^2$ for the six HFF clusters (Table  \ref{tab:numbers_highz}).  Based on these figures, we expect to find multiple images of high-redshift sources across almost the entire solid angle (91\%) of the WFC3 pointing on MACS\,J0717. Again, Abell 370 and Abell S1063 show the second- and third-highest fraction of WFC3 solid angle conducive to multiple-image creation. The lower values for Abell 2744, MACS\,J0416 and MACS\,J1149 are caused by their higher elongation and orientation within the WFC3 footprint on the sky.

\begin{figure}
\includegraphics[width=0.45\textwidth]{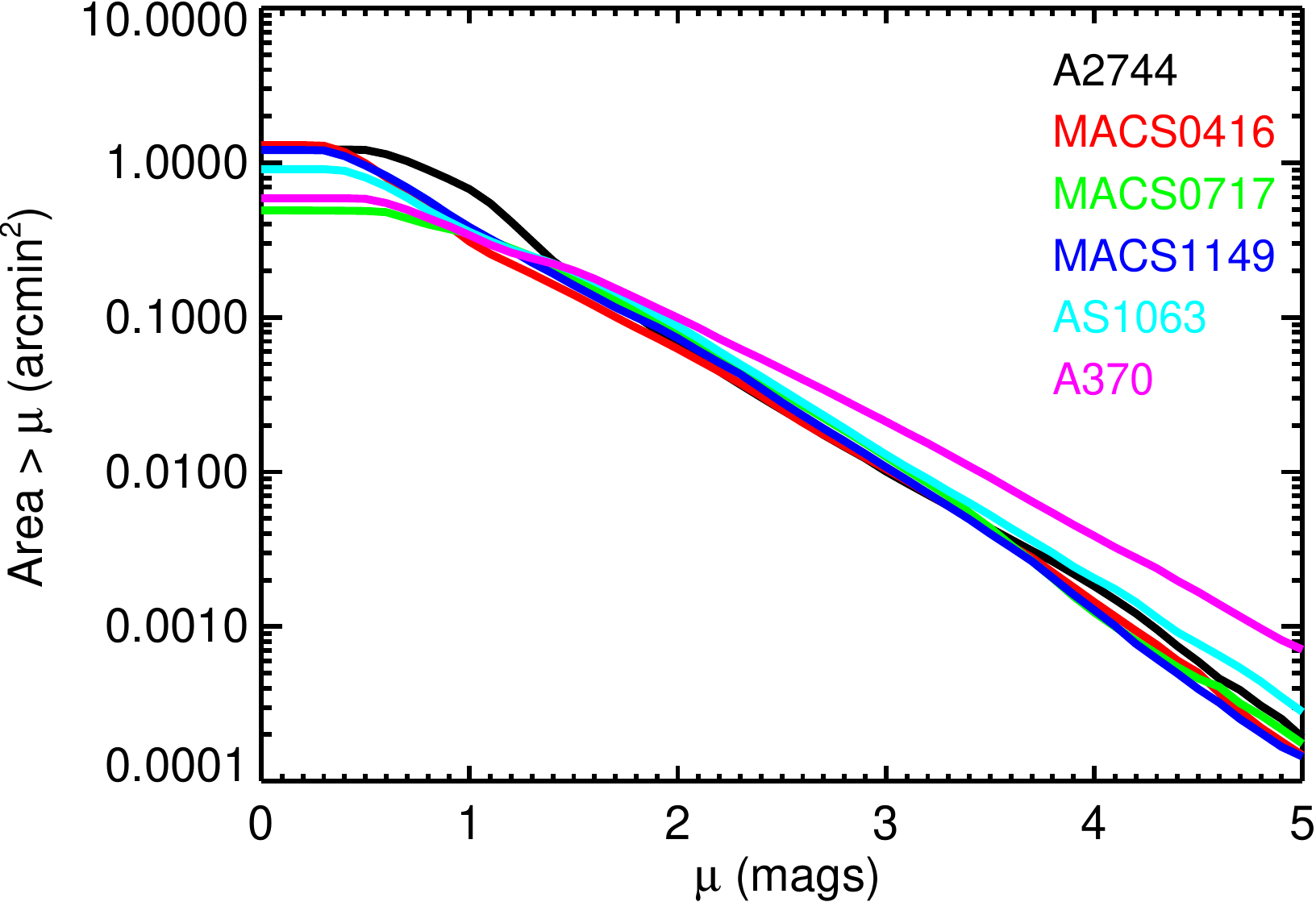}
\caption{\label{sourcesurface}Surface area in the source plane covered by the ACS+WFC3 region (see Figure \ref{sourcemag}) at a magnification above a given threshold $\mu$.}
\end{figure}

\begin{figure*}
\includegraphics[width=0.95\textwidth]{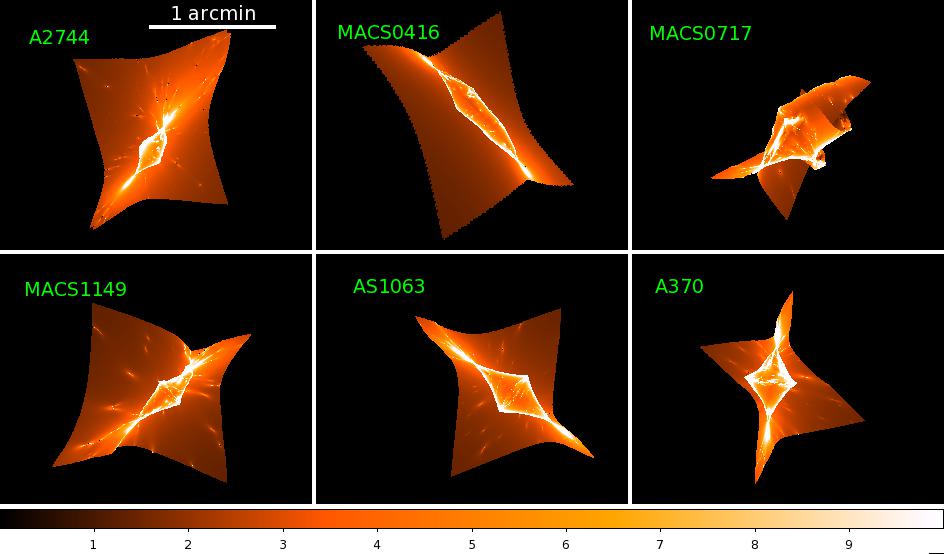}
\caption{\label{sourcemag} Source-plane magnification maps corresponding to the 
expected ACS+WFC3 coverage of the Frontier Field clusters. The color scale gives the magnification value.}
\end{figure*}

The combined effect of magnification and image multiplicity is best assessed in the source plane. To this end, we take advantage of Lenstool's capability to provide source-plane magnification maps based on, for a given source position, the most magnified image. Inverting the combined ACS+WFC3 aperture yields the source-plane magnification maps shown in Figure \ref{sourcemag}, where the regions of highest magnification now clearly delineate the \textit{caustic lines}. The strong variation in shape and surface area of these maps directly reflects the fraction of the respective HFF field that falls within the critical line. Indeed the total surface area in the source plane above a given magnification factor is directly proportional to the unlensed comoving volume covered at high redshift with this magnification (Figure \ref{sourcesurface}). As a result, \citet{2012ApJ...752..104W} proposed to use $\sigma_\mu$, the total surface area in the source plane above $\mu=3$, as a measurement of the efficiency of the lensing configuration to magnify high-redshift galaxies. These values are also reported in Table  \ref{tab:numbers_highz}.

\begin{table}
\begin{tabular}{llllll}
Cluster & $\Omega_i(\mu>5)$ & $\Omega_i(\mu>10)$ & $\Omega_i({\rm mult})$ & $\sigma_\mu(\mu>3)$ \\ 
&(\%) & (\%) & (\%)  & arcmin$^2$ \\
 \hline
Abell 2744   &  36  &  18  & 32 & 0.41 \\
MACS\,J0416  &  32  &  17  & 46 & 0.22 \\
MACS\,J0717  &  48  &  28  & 90 & 0.28 \\
MACS\,J1149  &  38  &  20  & 39 & 0.28 \\
Abell 370    &  61  &  38  & 65 & 0.28 \\
Abell S1063  &  47  &  25  & 55 & 0.26 \\
\end{tabular}
\caption{
\label{tab:numbers_highz} Lensing efficiency of the Hubble Frontier Field clusters using various metrics. From left to right: fraction of the area in the image plane amplified by $\mu>5$, by $\mu>10$, and covered by multiple images, and the surface area in the source plane magnified by $\mu>3$. All areas relate to the overlapping region with foreseen deep ACS and WFC3 observations. }
\end{table}

\begin{figure}
\includegraphics[width=0.5\textwidth]{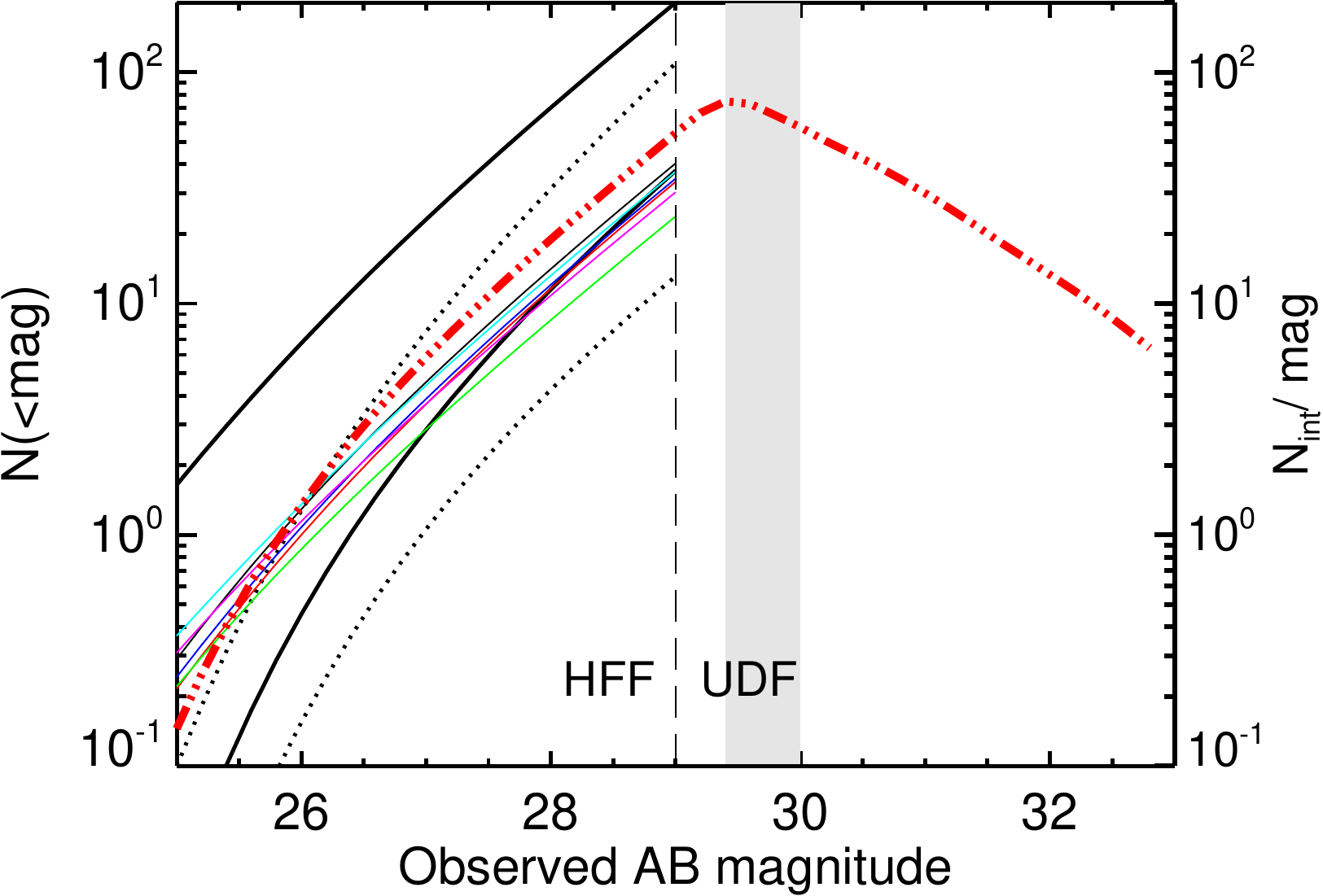}
\includegraphics[width=0.5\textwidth]{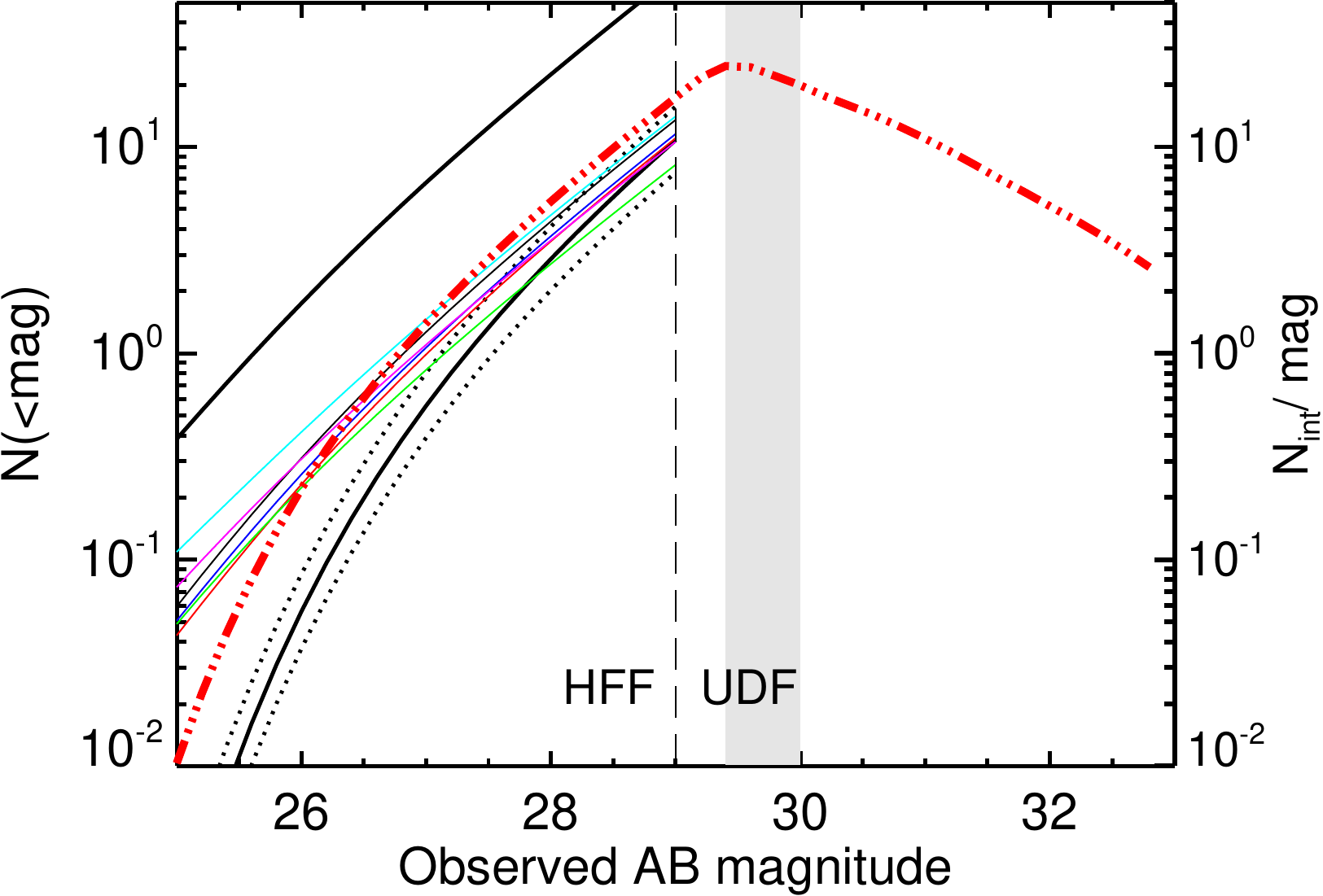}
\includegraphics[width=0.5\textwidth]{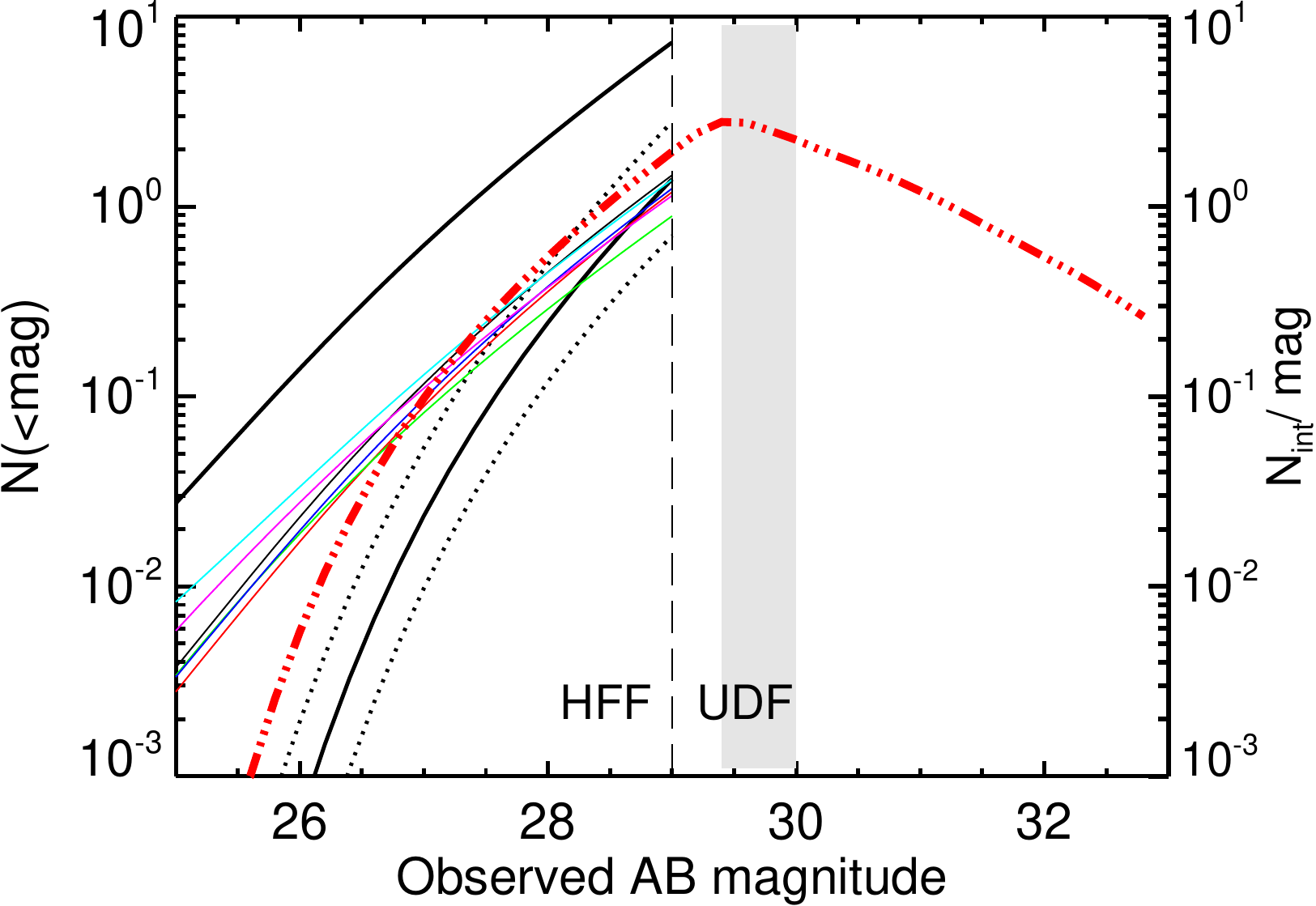}
\caption{\label{efficiency}Predicted number of high redshift sources 
at $z=7$ (top), $z=9$ (middle) and $z=11$ (bottom panel), per redshift 
interval $\Delta\,z=1$, assuming a blank field with typical errors 
(bottom black solid and dotted lines), each of the six FF clusters in turn (
same colours as Figure \ref{sourcemag}), and the sum over all fields (top solid black line).
Overplotted as a solid-dotted line in each panel are the total intrinsic counts (per magnitude bin, 
corrected for magnification) expected in the FF observations. The dashed limit and the gray region 
mark the limiting magnitudes for HFF and the UDF, respectively.
}
\end{figure}

Finally, we use the recent estimates on the UV luminosity function at high redshift from \citet{2014arXiv1403.4295B} to predict the number of high-redshift dropouts at $z\sim7$, $z\sim9$, and $z\sim11$ expected to be detected in the FF data, as a function of their observed (lensed) magnitude (Figure \ref{efficiency}). We assume an unbiased selection over a redshift interval $\Delta\,z$=1 centred on each redshift. The predictions at $z\sim9$ and $z\sim11$ use the fitting formula given by  \citet{2014arXiv1403.4295B} for the evolution of the \citet{Schechter76} parameters of the luminosity function at high redshift.  

The expected number counts at $z\sim7$ demonstrate the advantage provided by cluster lenses compared to a blank field of the same sky coverage.  The significant increase of bright sources thanks to gravitational lensing causes a \textit{positive magnification bias} at observed AB magnitudes $<27$, owing to the steep slope of the bright end of the UV luminosity function \citep{Maizy}.  This bright-end boost in the number counts exceeds a factor of 3 at mag$<26$, which is the typical limit for spectroscopic follow-up with current 8--10m class telescopes. The effect is even stronger at $z\sim9$ and $z\sim11$ (Figure \ref{efficiency}). 

The above predictions have already been tested by the deep near-infrared observations of the HFF cluster A2744, performed at the end of 2013.  Although ACS observations of matching depth are still lacking, the first searches for high-redshift  galaxies behind this cluster have independently identified 15 dropouts at $z\sim6-7$ down to an AB magnitude  of $J=28$ \citep{2014ApJ...786...60A}, and 18 dropouts at $z>7$ down to an AB magnitude $H=29$ \citep{2014arXiv1402.6743Z,2014A&A...562L...8L}. These numbers are slightly higher than the predictions for $z\sim7$ and $z\sim9$ dropouts in this cluster ($\sim$10 sources per $\Delta\,z=1$), due to either contamination of these dropout samples by lower-redshift sources (including low-mass stars), or most likely cosmic variance between cluster fields.

In summary, we expect a total of $\sim200$ $z=7$ dropouts, $\sim70$ $z=9$ dropouts and 5--10 dropouts at $z=11$ in the six Hubble Frontier Fields, per redshift interval $\Delta\,z=1$, and down to an observed AB magnitude of 29. While the number of high-redshift galaxies detected to this magnitude limit is very similar to that found in blank fields, gravitational magnification, which reaches a factor of $\sim1-3$ magnitudes in the central  regions of the HFF images, is the only way to access dropouts at even fainter magnitudes, down to m(AB)=30--32, 2 magnitudes fainter than the limits of the Hubble Ultra Deep Field. It will also increase (by a factor of at least 3) the survey sensitivity for galaxies at the highest redshifts (up to $z\sim11$) at observed magnitudes m(AB)$<$27.

We hope that the results of our efforts to calibrate the six HFF cluster lenses, described in this paper and made available to the community via the HFF website, will prove useful for the quantitative scientific exploitation of the HFF initiative and our quest to unravel the mysteries of the epoch of re-ionisation.

\section*{Acknowledgments}
JR acknowledges support from the ERC starting grant CALENDS and the CIG grant 294074.
MJ, ML and EJ acknowledge the M\'esocentre d'Aix-Marseille Universit\'e (project number : 14b030). 
This has also benefited from the facilities offered by CeSAM 
(Centre de donneS Astrophysique de Marseille\footnote{\tt http://lam.oamp.fr/cesam/}.
ML acknowledges the Centre National de la Recherche Scientifique (CNRS) for its support.
JPK and HA acknowledge support from the ERC advanced grant LIDA.
This work was supported by the Leverhulme Trust (grant number PLP-2011-003) and Science and Technology Facilities Council (grant number ST/F001166/1).

\bibliography{references}

\appendix
\section{Archival \textit{HST} observations of the HFF}
\begin{table*}
\label{tab:hst-obs}
\begin{tabular}{lcccccc}
Target & R.A. (J2000) Dec & Instrument & Filter & $t_{\rm obs}$ & Dataset & Obs.\ Date \\ \hline 
A2744-South & 00 14 18.7 \,\,\,$-$30 23 34 & ACS & F435W & 2725  & JB5G08010& 2009-10-30\\ 
A2744-South & 00 14 18.6 \,\,\,$-$30 23 37 & ACS & F435W & 5356  & JB5G06010& 2009-10-30\\ 
A2744-South & 00 14 18.6 \,\,\,$-$30 23 37 & ACS & F606W & 5356  & JB5G04010& 2009-10-29\\ 
A2744-South & 00 14 18.7 \,\,\,$-$30 23 34 & ACS & F606W & 1269  & JB5G08YYQ& 2009-10-30\\ 
A2744-South & 00 14 18.6 \,\,\,$-$30 23 37 & ACS & F814W & 5356  & JB5G02010& 2009-10-27\\ 
A2744-South & 00 14 18.7 \,\,\,$-$30 23 34 & ACS & F814W & 1268  & JB5G08YXQ& 2009-10-30\\ 
A2744-North & 00 14 22.2 \,\,\,$-$30 22 33 & ACS & F435W & 2725  & JB5G07010& 2009-10-30\\ 
A2744-North & 00 14 22.1 \,\,\,$-$30 22 36 & ACS & F435W & 5356  & JB5G05010& 2009-10-29\\ 
A2744-North & 00 14 22.1 \,\,\,$-$30 22 36 & ACS & F606W & 5356  & JB5G03010& 2009-10-27\\ 
A2744-North & 00 14 22.2 \,\,\,$-$30 22 33 & ACS & F606W & 1269  & JB5G07YQQ& 2009-10-30\\ 
A2744-North & 00 14 22.2 \,\,\,$-$30 22 33 & ACS & F814W & 1268  & JB5G07YPQ& 2009-10-30\\ 
A2744-North & 00 14 22.1 \,\,\,$-$30 22 36 & ACS & F814W & 5356  & JB5G01010& 2009-10-27\\ \hline
MACSJ0416-2403 &   04 16 08.4 \,\,\,$-$24 04 21  &  WFC3&  F225W& 3634 & IBSTA6030   &2012-08-18   \\
MACSJ0416-2403 &   04 16 08.4 \,\,\,$-$24 04 21  &  WFC3& F275W & 3684 &  IBSTB3040  & 2012-09-02  \\
MACSJ0416-2403 &   04 16 08.4 \,\,\,$-$24 04 21  &  WFC3&  F336W& 2360 & IBSTB6030   &2012-09-14   \\
MACSJ0416-2403 &   04 16 08.4 \,\,\,$-$24 04 21  &  WFC3&  F390W& 1156 & IBSTB3030   &2012-09-02   \\
MACSJ0416-2403 &   04 16 08.4 \,\,\,$-$24 04 21  &  WFC3& F390W & 1251 &  IBSTA6040  & 2012-08-18  \\
MACSJ0416-2403 &   04 16 08.4 \,\,\,$-$24 04 20  &  ACS &   F435W& 1020&   JBSTB5020 &  2012-09-14 \\
MACSJ0416-2403 &   04 16 08.4 \,\,\,$-$24 04 21  &  ACS &   F435W& 1032&   JBSTB0010 &  2012-08-20 \\
MACSJ0416-2403 &   04 16 08.4 \,\,\,$-$24 04 21  &  ACS &   F475W& 1032&   JBSTA1010 &  2012-07-24 \\
MACSJ0416-2403 &   04 16 08.4 \,\,\,$-$24 04 21  &  ACS &   F475W& 1032&   JBSTB7010 &  2012-09-27 \\
MACSJ0416-2403 &   04 16 08.4 \,\,\,$-$24 04 20  &  ACS &   F606W&   986&   JBSTA8020&   2012-08-31\\
MACSJ0416-2403 &   04 16 08.4 \,\,\,$-$24 04 21  &  ACS &   F606W& 1032&   JBSTA3010 &  2012-08-05 \\
MACSJ0416-2403 &   04 16 08.4 \,\,\,$-$24 04 21  &  ACS &   F625W&   985&    JBSTB0020&     2012-08-20\\
MACSJ0416-2403 &   04 16 08.4 \,\,\,$-$24 04 21  &  ACS &   F625W& 1032&   JBSTA0010 &  2012-07-24 \\
MACSJ0416-2403 &   04 16 08.4 \,\,\,$-$24 04 20  &  ACS &   F775W& 1016&   JBSTB2020 &  2012-09-02 \\
MACSJ0416-2403 &   04 16 08.4 \,\,\,$-$24 04 21  &  ACS &   F775W&  1015&   JBSTA1020&   2012-07-24\\
MACSJ0416-2403 &   04 16 08.4 \,\,\,$-$24 04 20  &  ACS &   F814W&   986&   JBSTA3020&   2012-08-05\\
MACSJ0416-2403 &   04 16 08.4 \,\,\,$-$24 04 20  &  ACS &   F814W&   987&   JBSTA5020&   2012-08-18\\
MACSJ0416-2403 &   04 16 08.4 \,\,\,$-$24 04 21  &  ACS &   F814W& 1032&   JBSTA8010 &  2012-08-31 \\
MACSJ0416-2403 &   04 16 08.4 \,\,\,$-$24 04 21  &  ACS &   F814W& 1032&   JBSTB5010 &  2012-09-14 \\
MACSJ0416-2403 &   04 16 08.4 \,\,\,$-$24 04 21  &  ACS &   F850LP&  1019&   JBSTA0020&   2012-07-24\\
MACSJ0416-2403 &   04 16 08.4 \,\,\,$-$24 04 20  &  ACS &   F850LP& 1003&   JBSTB7020 &  2012-09-27 \\
MACSJ0416-2403 &   04 16 08.4 \,\,\,$-$24 04 21  &  ACS &   F850LP& 1032&   JBSTA5010 &  2012-08-18 \\
MACSJ0416-2403 &   04 16 08.4 \,\,\,$-$24 04 21  &  ACS &   F850LP& 1032&   JBSTB2010 &  2012-09-02 \\
MACSJ0416-2403 &   04 16 08.4 \,\,\,$-$24 04 20  &  WFC3&  F105W& 1305 &  IBSTB6050  & 2012-09-14  \\
MACSJ0416-2403 &   04 16 08.4 \,\,\,$-$24 04 21  &  WFC3&  F105W& 1509 & IBSTA4020   &2012-08-05   \\
MACSJ0416-2403 &   04 16 08.4 \,\,\,$-$24 04 21  &  WFC3&  F110W& 1509 & IBSTA2020   &2012-07-24   \\
MACSJ0416-2403 &   04 16 08.4 \,\,\,$-$24 04 21  &  WFC3& F110W & 1006 &  IBSTA9030  & 2012-08-31  \\
MACSJ0416-2403 &   04 16 08.4 \,\,\,$-$24 04 21  &  WFC3&  F125W& 1509 & IBSTB1020   &2012-08-20   \\
MACSJ0416-2403 &   04 16 08.4 \,\,\,$-$24 04 21  &  WFC3& F125W & 1006 &  IBSTB8030  & 2012-09-27  \\ 
MACSJ0416-2403 &   04 16 08.4 \,\,\,$-$24 04 20  &  WFC3&  F140W& 1005 &  IBSTB1030  & 2012-08-20  \\
MACSJ0416-2403 &   04 16 08.4 \,\,\,$-$24 04 21  &  WFC3&  F140W& 1306 & IBSTB6040   &2012-09-14   \\
MACSJ0416-2403 &   04 16 08.4 \,\,\,$-$24 04 20  &  WFC3&  F160W& 1005 &  IBSTA2030  & 2012-07-24  \\
MACSJ0416-2403 &   04 16 08.4 \,\,\,$-$24 04 21  &  WFC3&  F160W& 1509 & IBSTA9020   &2012-08-31   \\
MACSJ0416-2403 &   04 16 08.4 \,\,\,$-$24 04 21  &  WFC3&  F160W& 1509 & IBSTB8020   &2012-09-27   \\
MACSJ0416-2403 &   04 16 08.4 \,\,\,$-$24 04 21  &  WFC3& F160W & 1006 &  IBSTA4030  & 2012-08-05  \\ \hline
MACS0717+3745         & 07 17 32.6 \,\,\,$+$37 45 00 & WFC3 & F225W & 3645 &BFLA6030 &2011-11-19 \\
MACS0717+3745         & 07 17 32.6 \,\,\,$+$37 45 00 & WFC3 & F275W & 3723 &BFLB3040 &2011-09-20 \\
MACS0717+3745         & 07 17 32.6 \,\,\,$+$37 45 00 & WFC3 & F336W & 2391 &BFLB6030 &2011-10-10 \\
MACS0717+3745         & 07 17 32.6 \,\,\,$+$37 45 00 & WFC3 & F390W & 1254 &BFLA6040 &2011-11-19\\
MACS0717+3745         & 07 17 32.6 \,\,\,$+$37 45 00 & WFC3 & F390W & 1179 &BFLB3030 &2011-09-20\\
MACS0717+3745         & 07 17 32.6 \,\,\,$+$37 45 00 & ACS  &  F435W & 1032 &JBFLA3010 &2011-10-29\\
MACS0717+3745         & 07 17 32.7 \,\,\,$+$37 45 01 & ACS  &  F435W & 994  &JBFLB5020 &2011-10-10 \\
MACS0717+3745         & 07 17 32.7 \,\,\,$+$37 45 01 & ACS  &  F475W &1000 &JBFLA3020 &2011-10-29\\
MACS0717+3745         & 07 17 32.6 \,\,\,$+$37 45 00 & ACS  &  F475W &1032 &JBFLB7010 &2011-10-30\\
MACSJ0717+3745        & 07 17 32.9 \,\,\,$+$37 45 05 & ACS  &  F555W &4470 &J8QU05010 &2004-04-02\\
\end{tabular}
\caption{Archival \textit{HST} imaging observations (ACS and WFC3) broad-band filters only) of the HFF as of July 2013. The fields are listed in the planned order of observation.}
\end{table*}

\setcounter{table}{0}
\begin{table*}
\begin{tabular}{lcccccc}
MACSJ0717.5+3745-POS5 & 07 17 32.0 \,\,\,$+$37 44 48 & ACS  &  F606W &1980 &J97001010 &2005-01-25\\
MACSJ0717.5+3745-POS5 & 07 17 43.4 \,\,\,$+$37 47 01 & ACS  &  F606W &1980 &J97005010 &2005-01-30\\
MACS0717+3745         & 07 17 32.6 $+$37 45 00 & ACS    &F625W & 1032 &JBFLA0010 &2011-10-10\\
MACS0717+3745         & 07 17 32.6 $+$37 45 00 & ACS    &F625W & 1032 &JBFLB5010 &2011-10-10\\
MACS0717+3745         & 07 17 32.7 $+$37 45 01 & ACS    &F775W & 1023 &JBFLA8020 &2011-12-08\\
MACS0717+3745         & 07 17 32.6 $+$37 45 00 & ACS    &F775W & 1023 &JBFLB0020 &2011-08-31\\
MACSJ0717+3745        & 07 17 32.9 $+$37 45 05 & ACS    &F814W & 2097 &J9OI04010 &2006-10-13\\
MACSJ0717+3745        & 07 17 32.9 $+$37 45 05 & ACS    &F814W & 4560 &J8QU05020 &2004-04-02\\
MACSJ0717.5+3745-POS5 & 07 17 43.4 \,\,\,$+$37 47 01 & ACS   & F814W & 4020 &J97005020 &2005-01-30\\
MACSJ0717+3745        & 07 17 32.9 \,\,\,$+$37 45 05 & ACS & F814W  &2236 &J9DD04010 &2005-10-25 \\
MACS0717+3745         & 07 17 32.6 \,\,\,$+$37 45 00 & ACS & F850LP & 1026 &JBFLA0020 &2011-10-10 \\
MACS0717+3745         & 07 17 32.6 \,\,\,$+$37 45 00 & ACS & F850LP & 1032 &JBFLA8010 &2011-12-08 \\
MACS0717+3745         & 07 17 32.6 \,\,\,$+$37 45 00 & ACS & F850LP & 1032 &JBFLB0010 &2011-08-31 \\
MACS0717+3745         & 07 17 32.7 \,\,\,$+$37 45 01 & ACS & F850LP & 1010 &JBFLB7020 &2011-10-30 \\
MACS0717+3745         & 07 17 32.6 \,\,\,$+$37 45 00 & WFC3& F105W &1509 &IBFLA4020 &2011-10-29\\
MACS0717+3745         & 07 17 32.7 \,\,\,$+$37 45 00 & WFC3& F105W &1306 &IBFLB6050 &2011-10-11\\
MACS0717+3745         & 07 17 32.6 \,\,\,$+$37 45 00 & WFC3& F110W &1509 &IBFLA2020 &2011-10-10\\
MACS0717+3745         & 07 17 32.6 \,\,\,$+$37 44 59 & WFC3& F110W &1006 &IBFLA9030 &2011-12-09\\
MACS0717+3745         & 07 17 32.6 \,\,\,$+$37 45 00 & WFC3& F125W &1509 &IBFLB1020 &2011-08-31\\
MACS0717+3745         & 07 17 32.6 \,\,\,$+$37 44 59 & WFC3& F125W &1006 &IBFLB8030 &2011-10-30\\
MACS0717+3745         & 07 17 32.6 \,\,\,$+$37 44 59 & WFC3& F140W &1006 &IBFLA4030 &2011-10-29\\
MACS0717+3745         & 07 17 32.6 \,\,\,$+$37 45 00 & WFC3& F140W &1306 &IBFLB6040 &2011-10-11\\
MACS0717+3745         & 07 17 32.6 \,\,\,$+$37 45 00 & WFC3& F160W &1006 &IBFLA2030 &2011-10-10\\
MACS0717+3745         & 07 17 32.6 \,\,\,$+$37 45 00 & WFC3& F160W &1509 &IBFLA9020 &2011-12-09\\
MACS0717+3745         & 07 17 32.6 \,\,\,$+$37 45 00 & WFC3& F160W &1006 &IBFLB1030 &2011-09-01\\
MACS0717+3745         & 07 17 32.6 \,\,\,$+$37 45 00 & WFC3& F160W &1509 &IBFLB8020 &2011-10-30\\ \hline
MACS1149+2223  & 11 49 35.7 $+$22 23 55 & WFC3  &  F225W & 1194 &  IBF5A6020 &  2011-02-13 \\ 
MACS1149+2223  & 11 49 35.7 $+$22 23 55 & WFC3  &  F225W & 2362 &  IBF5B6030 &  2011-02-27 \\
MACS1149+2223  & 11 49 35.7 $+$22 23 55 & WFC3  &  F275W & 1194 &  IBF5A6030 &  2011-02-13 \\
MACS1149+2223  & 11 49 35.7 $+$22 23 55 & WFC3  &  F275W & 2414 &  IBF5B4020 &  2011-02-15 \\
MACS1149+2223  & 11 49 35.7 $+$22 23 55 & WFC3  &  F336W & 1195 &  IBF5A7020 &  2011-02-13 \\
MACS1149+2223  & 11 49 35.7 $+$22 23 54 & WFC3  &  F336W & 1196 &  IBF5B3030 &  2011-02-15 \\
MACS1149+2223  & 11 49 35.7 $+$22 23 55 & WFC3  &  F390W & 1196 &  IBF5A7030 &  2011-02-13 \\
MACS1149+2223  & 11 49 35.7 $+$22 23 55 & WFC3  &  F390W & 1195 &  IBF5B3020 &  2011-02-15 \\
MACS1149+2223  & 11 49 35.7 $+$22 23 55 & ACS   &  F435W &  1032 &  JBF5A5010&   2011-02-13\\
MACS1149+2223  & 11 49 35.7 $+$22 23 56 & ACS   &  F435W &   956 &   JBF5B5020 &  2011-02-27\\
MACS1149+2223  & 11 49 35.7 $+$22 23 56 & ACS   &  F475W &  1034 &  JBF5A3020&   2011-01-30\\
MACS1149+2223  & 11 49 35.7 $+$22 23 55 & ACS   &  F475W &  1034 &  JBF5B0020&   2010-12-04\\
MACSJ1149+2223 & 11 49 35.5 $+$22 24 04 & ACS   &  F555W &  4500 &  J8QU08010&  2004-04-22 \\
MACS1149+2223  & 11 49 35.7 $+$22 23 55 & ACS   &  F606W &  1032 &  JBF5A0010&   2011-01-15\\
MACS1149+2223  & 11 49 35.7 $+$22 23 55 & ACS   &  F606W &  1032 &  JBF5B5010&   2011-02-27\\
MACS1149+2223  & 11 49 35.7 $+$22 23 56 & ACS   &  F625W &  1015 &  JBF5A5020&   2011-02-13\\
MACS1149+2223  & 11 49 35.7 $+$22 23 55 & ACS   &  F625W &  1032 &  JBF5B7010&   2011-03-09\\
MACS1149+2223  & 11 49 35.7 $+$22 23 56 & ACS   &  F775W &   994  &  JBF5A8020&   2011-02-27\\
MACS1149+2223  & 11 49 35.7 $+$22 23 56 & ACS   &  F775W &  1053 &  JBF5B7020&   2011-03-09\\
MACSJ1149+2223 & 11 49 35.5 $+$22 24 04 & ACS   &  F814W &  4590 &  J8QU08020&  2004-04-22 \\
MACSJ1149+2223 & 11 49 35.5 $+$22 24 04 & ACS   &  F814W &  2184 &  J9DD07010&  2006-05-25 \\
MACS1149+2223  & 11 49 35.7 $+$22 23 55 & ACS   &  F850LP&  1044 &  JBF5A0020&   2011-01-15\\
MACS1149+2223  & 11 49 35.7 $+$22 23 55 & ACS   &  F850LP&  1032 &  JBF5A3010&   2011-01-30\\
MACS1149+2223  & 11 49 35.7 $+$22 23 55 & ACS   &  F850LP&  1032 &  JBF5A8010&  2011-02-27 \\
MACS1149+2223  & 11 49 35.7 $+$22 23 55 & ACS   &  F850LP&  1032 &  JBF5B0010&   2010-12-04\\
MACS1149+2223  & 11 49 35.7 $+$22 23 55 & WFC3  &  F105W & 1509 &  IBF5A4020 &  2011-01-30 \\
MACS1149+2223  & 11 49 35.7 $+$22 23 55 & WFC3  &  F105W & 1306 &  IBF5B6050 &  2011-02-27 \\
MACS1149+2223  & 11 49 35.7 $+$22 23 55 & WFC3  &  F110W & 1509 &  IBF5A2020 &  2011-01-16 \\
MACS1149+2223  & 11 49 35.7 $+$22 23 54 & WFC3  &  F110W & 906  &   IBF5A9030&   2011-02-27\\
MACS1149+2223  & 11 49 35.7 $+$22 23 55 & WFC3  &  F125W & 1509 &  IBF5B1020 &  2010-12-04 \\
MACS1149+2223  & 11 49 35.7 $+$22 23 54 & WFC3  &  F125W & 1006 &  IBF5B8030 &  2011-03-09 \\
MACS1149+2223  & 11 49 35.7 $+$22 23 54 & WFC3  &  F140W & 1006 &  IBF5A4030 &  2011-01-30 \\
MACS1149+2223  & 11 49 35.7 $+$22 23 55 & WFC3  &  F140W & 1306 &  IBF5B6040 &  2011-02-27 \\
MACS1149+2223  & 11 49 35.7 $+$22 23 55 & WFC3  &  F160W & 1006 &  IBF5A2030 &  2011-01-16 \\
MACS1149+2223  & 11 49 35.7 $+$22 23 55 & WFC3  &  F160W & 1509 &  IBF5A9020 &  2011-02-27 \\
MACS1149+2223  & 11 49 35.7 $+$22 23 55 & WFC3  &  F160W & 1005 &  IBF5B1030 &  2010-12-04 \\
MACS1149+2223  & 11 49 35.7 $+$22 23 55 & WFC3  &  F160W & 1509 &  IBF5B8020 &  2011-03-09 \\ \hline
\end{tabular}
\caption{\emph{(continued)}}   
\end{table*}

\setcounter{table}{0}
\begin{table*}
\begin{tabular}{lcccccc}
Target & R.A. (J2000) Dec & Instrument & Filter & $t_{\rm obs}$ & Dataset & Obs.\ Date \\ \hline 
RXJ2248-4431    &22 48 44.0  \,\,\,$-$44 31 51 &   WFC3  &        F225W   & 3574  &  BSUA6030   &    2012-09-24  \\
RXJ2248-4431    &22 48 44.0  \,\,\,$-$44 31 51 &   WFC3  &        F275W   & 3637  &  BSUB3040   &    2012-10-09  \\
RXJ2248-4431    &22 48 44.0  \,\,\,$-$44 31 51 &   WFC3  &        F336W   & 2359  &  BSUB6030   &    2012-10-22  \\
RXJ2248-4431    &22 48 44.0  \,\,\,$-$44 31 51 &   WFC3  &        F390W   & 1215  &  BSUA6040   &    2012-09-24  \\
RXJ2248-4431    &22 48 44.0  \,\,\,$-$44 31 51 &   WFC3  &        F390W   & 1155  &  BSUB3030   &    2012-10-09  \\
RXJ2248-4431    &22 48 44.0  \,\,\,$-$44 31 51 &   ACS     &        F435W   & 1032  &  JBSUA8010&       2012-10-04\\
RXJ2248-4431    &22 48 44.0  \,\,\,$-$44 31 52 &   ACS     &        F435W   & 1019  &  JBSUB5020&       2012-10-22\\
RXJ2248-4431    &22 48 44.0  \,\,\,$-$44 31 51 &   ACS     &        F475W   & 1032  &  JBSUA1010&       2012-08-30\\
RXJ2248-4431    &22 48 44.0  \,\,\,$-$44 31 51 &   ACS     &        F475W   & 1032  &  JBSUB7010&       2012-11-04\\
RXJ2248-4431    &22 48 44.0  \,\,\,$-$44 31 52 &   ACS     &        F606W   &    985 &  JBSUA3020&       2012-09-12\\
RXJ2248-4431    &22 48 44.0  \,\,\,$-$44 31 51 &   ACS     &        F606W   &  1003 &  JBSUB0020&       2012-09-26\\
RXJ2248-4431    &22 48 44.0  \,\,\,$-$44 31 51 &   ACS     &        F625W   &  1032 &  JBSUA0010&       2012-08-30\\
RXJ2248-4431    &22 48 44.0  \,\,\,$-$44 31 51 &   ACS     &        F625W   &  1032 &  JBSUB0010&       2012-09-26\\
RXJ2248-4431    &22 48 44.0  \,\,\,$-$44 31 51 &   ACS     &        F775W   &  1014 &  JBSUA1020&       2012-08-30\\
RXJ2248-4431    &22 48 44.0  \,\,\,$-$44 31 52 &   ACS     &        F775W   &  1015 &  JBSUB2020&       2012-10-09\\
RXJ2248-4431    &22 48 44.0  \,\,\,$-$44 31 51 &   ACS     &        F814W   &   1032&  JBSUA3010&       2012-09-12\\
RXJ2248-4431    &22 48 44.0  \,\,\,$-$44 31 52 &   ACS     &        F814W   &  986   &  JBSUA5020&       2012-09-24\\
RXJ2248-4431    &22 48 44.0  \,\,\,$-$44 31 52 &   ACS     &        F814W   &  1018 &  JBSUA8020&       2012-10-04\\
RXJ2248-4431    &22 48 44.0  \,\,\,$-$44 31 51 &   ACS     &        F814W   &  1032 &  JBSUB5010&       2012-10-22\\
ANY                       &22 48 44.7  \,\,\,$-$44 31 38 &  ACS      &        F814W   &  1918 &  JC6HS1010  &     2012-11-19\\
RXJ2248-4431    &22 48 44.0  \,\,\,$-$44 31 51 &   ACS     &        F850LP &  1018 &  JBSUA0020 &      2012-08-30\\ 
RXJ2248-4431    &22 48 44.0  \,\,\,$-$44 31 51 &   ACS     &        F850LP &  1032 &  JBSUA5010 &      2012-09-24\\ 
RXJ2248-4431    &22 48 44.0  \,\,\,$-$44 31 51 &   ACS     &        F850LP &  1032 &  JBSUB2010&       2012-10-09\\
RXJ2248-4431    &22 48 44.0  \,\,\,$-$44 31 52 &   ACS     &        F850LP &  1002 &  JBSUB7020&       2012-11-04\\
RXJ2248-4431    &22 48 44.0  \,\,\,$-$44 31 51 &   WFC3  &        F105W  &  1509 &  IBSUA4020 &      2012-09-12\\
RXJ2248-4431    &22 48 43.9  \,\,\,$-$44 31 52 &   WFC3  &        F105W  &  1306 &  IBSUB6050 &      2012-10-22\\
RXJ2248-4431    &22 48 44.0  \,\,\,$-$44 31 51 &   WFC3  &        F110W  &  1509 &  IBSUA2020 &      2012-08-30\\
RXJ2248-4431    &22 48 43.9  \,\,\,$-$44 31 51 &   WFC3  &        F110W  &  1006 &  IBSUA9030 &      2012-10-04\\
RXJ2248-4431    &22 48 44.0  \,\,\,$-$44 31 51 &   WFC3  &        F125W  &  1509 &  IBSUB1020 &      2012-09-26\\
RXJ2248-4431    &22 48 43.9  \,\,\,$-$44 31 51 &   WFC3  &        F125W  &  1006 &  IBSUB8030 &      2012-11-04\\
RXJ2248-4431    &22 48 43.9  \,\,\,$-$44 31 51 &   WFC3  &        F140W  &  1006 &  IBSUA4030 &      2012-09-12\\
RXJ2248-4431    &22 48 44.0  \,\,\,$-$44 31 51 &   WFC3  &        F140W  &  1306 &  IBSUB6040 &      2012-10-22\\
RXJ2248-4431    &22 48 44.0  \,\,\,$-$44 31 51 &   WFC3  &        F160W  &  1006 &  IBSUA2030 &      2012-08-30\\
RXJ2248-4431    &22 48 44.0  \,\,\,$-$44 31 51 &   WFC3  &        F160W  &  1509 &  IBSUA9020 &      2012-10-04\\
RXJ2248-4431    &22 48 44.0  \,\,\,$-$44 31 52 &   WFC3  &        F160W  &  1006 &  IBSUB1030 &      2012-09-26\\
RXJ2248-4431    &22 48 44.0  \,\,\,$-$44 31 51 &   WFC3  &        F160W  &  1509 &  IBSUB8020 &      2012-11-04\\ \hline
ABELL-0370                 & 02 39 51.0 \,\,\,$-$01 34 50 & ACS & F475W & 6780 & JABU01030 & 2009-07-16 \\
SMMJ02399$-$0136  & 02 39 52.0 \,\,\,$-$01 35 58 & ACS & F475W & 2250 & JB3402011  & 2010-12-25 \\
ABELL-0370                 & 02 39 51.0 \,\,\,$-$01 34 50 & ACS & F625W & 2040 & JABU01010 & 2009-07-16 \\
ABELL-0370                 & 02 39 51.0 \,\,\,$-$01 34 50 & ACS & F814W & 3840 & JABU01020 & 2009-07-16 \\
ABELL-370                   & 02 39 51.5 \,\,\,$-$01 34 46 & ACS & F814W & 4720 & JB5M22010 & 2010-12-20\\ 
ABELL-370                   & 02 39 51.5 \,\,\,$-$01 34 48 & ACS & F814W & 4880 & JB5M22020 & 2010-12-20 \\
ABELL-370-WFC3      & 02 39 53.9 \,\,\,$-$01 34 32 & WFC3 & F110W & 2612 & IB5M12020 & 2010-12-19 \\
ABELL-370-WFC3      & 02 39 53.9 \,\,\,$-$01 34 32 & WFC3 & F160W & 2412 & IB5M12010 & 2010-12-19 \\ \hline
\end{tabular}
\caption{\emph{(continued)}}   
\end{table*}

\end{document}